\documentclass[12pt]{article}
\usepackage{epsfig}
\usepackage{amssymb}
\usepackage{amsmath}
\usepackage{amsfonts}
\usepackage{graphicx}
\usepackage{mathrsfs}
\usepackage[dvips]{color}
\usepackage{multirow}
\usepackage{subfig}
\usepackage{float}
\usepackage{epstopdf}

\newcommand{\R}{\mathbb{R}}
\newcommand{\C}{\mathbb{C}}

\newcommand{\ff}{\mathfrak{f}}
\newcommand{\fg}{\mathfrak{g}}

\newcommand{\ft}{\mathfrak{t}}

\newcommand{\fz}{\mathfrak{z}}

\newcommand{\fK}{\mathfrak{K}}

\newcommand{\bfe}{\mathbf{e}}

\newcommand{\bk}{\mathbf{k}}

\newcommand{\bp}{\mathbf{p}}

\newcommand{\bx}{\mathbf{x}}

\newcommand{\bA}{\mathbf{A}}

\newcommand{\cL}{\mathcal{L}}

\newcommand{\cO}{\mathcal{O}}

\newcommand{\be}{\begin{equation}}
\newcommand{\ee}{\end{equation}}
\newcommand{\bea}{\begin{eqnarray}}
\newcommand{\eea}{\end{eqnarray}}
\newcommand{\nn}{\nonumber}
\newcommand{\kt}{\rangle}
\newcommand{\br}{\langle}

\newcommand{\ed}{\end{document}}

\newcommand{\bi}{\begin{itemize}}
\newcommand{\ei}{\end{itemize}}

\newcommand{\bce}{\begin{center}}
\newcommand{\ece}{\end{center}}

\newcommand{\sG}{\mathscr{G}}
\newcommand{\sH}{\mathscr{H}}

\newcommand{\ket}[1]{\mid\!\! \, #1\rangle}                              %
\usepackage{mathtools}
\DeclarePairedDelimiterX\MeijerM[3]{\lparen}{\rparen}%
{\begin{smallmatrix}#1 \\ #2\end{smallmatrix}\delimsize\vert\,#3}

\begin{document}

\title{Geometric scattering in the presence of line defects}

\author{Hai Viet Bui\thanks{E-mail address: haibui@utexas.edu}, Ali~Mostafazadeh\thanks{E-mail address:
amostafazadeh@ku.edu.tr}, and Sema Seymen\thanks{E-mail address: seymen@itu.edu.tr}\\[6pt]
$^*$Department of Chemistry and Physics, Augusta University,\\
1120 15th Street, Augusta, GA 30912, US\\[6pt]
$^\dagger$Departments of Mathematics and Physics, Ko\c{c} University,\\  34450 Sar{\i}yer,
Istanbul, Turkey\\[6pt]
$^\ddagger$Department of Physics, Bo\u{g}azi\c{c}i University,\\ 34342 Bebek, Istanbul, Turkey}

\date{}
\maketitle

\begin{abstract}

A non-relativistic scalar particle moving on a curved surface
undergoes a geometric scattering whose behavior is sensitive to the
theoretically ambiguous values of the intrinsic and extrinsic
curvature coefficients entering the expression for the quantum
Hamiltonian operator. This suggests using the scattering data to
settle the ambiguity in the definition of the Hamiltonian. It has
recently been shown that the inclusion of point defects on the
surface enhances the geometric scattering effects. We perform a
detailed study of the geometric scattering phenomenon in the
presence of line defects for the case that the particle is confined
to move on a Gaussian bump and the defect(s) are modeled by
delta-function potentials supported on a line or a set of parallel
lines normal to the scattering axis. In contrast to a surface having
point defects, the scattering phenomenon associated with this system
is generically geometric in nature in the sense that for a flat
surface the scattering amplitude vanishes for all scattering angles
$\theta$ except $\theta=\theta_0$ and $\pi-\theta_0$, where
$\theta_0$ is the angle of incidence. We show that the presence of
the line defects amplifies the geometric scattering due to the
Gaussian bump. This amplification effect is particularly strong when
the center of the bump is placed between two line defects.

\vspace{2mm}



\end{abstract}

\section{Introduction}

Quantum mechanics in a curved space has a long history. As a first
step to formulate a quantum theory of gravity, it has attracted a
lot of attention during the the past seven decades
\cite{Dewitt1,Penrose,Dewitt2,Marinov-1980,Kleinert,Dewitt3}. Among
the basic difficulties in extending non-relativistic quantum
mechanics to a curved space is the ambiguity in the identification
of the Hamiltonian operator. This stems from the notorious
factor-ordering problem. If a free particle moves on a Riemannian
manifold with metric $\fg$, the classical Hamiltonian takes the
form, $H_c=\frac{1}{2m}g^{ij}(x)p_ip_j$, where $g^{ij}$ are the
entries of the inverse of the matrix formed out of the components
$g_{ij}$ of $\fg$ in a local coordinate chart, $p_i$ are the momenta
conjugate to the local coordinates $x_i$ of the points of $M$, and
$x:=(x^1,x^2,\cdots,x^{{\rm dim}(M)})$. Standard operator
quantization of this Hamiltonian together with the requirement that
the Hamiltonian operator must act as a self-adjoint operator in the
Hilbert space of square-integrable functions $\psi:M\to\C$ do not
fix it, because there are an infinity of admissible choices for the
ordering of the factors $g^{ij}(x)$, $p_i$, and $p_j$.

The assumption that the Hamiltonian operator $H$ must transform as a scalar under coordinate transformation reduces the ambiguity in the definition of the Hamiltonian operator to the freedom in the choice of a real coefficient $\lambda$ that enters the following expression for the Hamiltonian.
    \be
    H= -\frac{\hbar^2}{2m} g^{-1/2}\partial_i (g^{ij} g^{1/2})\partial_j +\frac{\lambda\hbar^2}{m} R.
    \label{H=}
    \ee
Here and in what follows, we employ Einstein's summation convention
for repeated indices, $g$ is the determinant of the matrix
$[g_{ij}]$, and $R$ is the Ricci scalar curvature associated with
the metric $\fg$. The path-integral quantization scheme does not
offer a remedy either, because in this scheme the factor-ordering
ambiguity manifests itself in the form of the non-uniqueness of the
path-integral measure.

Since theoretical considerations do not fix the value of $\lambda$,
one may view (\ref{H=}) with different choices for $\lambda$ as the
Hamiltonian operator for different quantum systems. The presence of
the curvature term in (\ref{H=}) does indeed contribute to
physically measurable quantities associated with the system
\cite{Ferrari-2008,Szamiet-2010,Della-Valle-2010,Jensen-2011,Silva-2013,Pahlavani-2015}.
For the cases that $M$ is compact and the Hamiltonian operator has a
discrete spectrum, it affects the transition energies. For the cases
where $M$ is an asymptotically flat manifold allowing for a
well-posed scattering problem, it contributes to the scattering
amplitude. In the latter case, one should in principle be able to
determine $\lambda$ by performing scattering experiments.

The simplest scenario is to consider a scattering setup in which a free particle moving on an asymptotically flat surface $S$ scatters due to the nontrivial geometry of $S$, \cite{pra-1996}. In practice, however, one can confine the particle to move on the surface, if there are confining forces that prevent its motion along the normal direction to the surface. Such a surface is clearly embedded in the Euclidean space $\R^3$, and one can model the effect of the confining forces in terms of the thin-layer quantization scheme of Ref.~\cite{daCosta-1981}. The resulting Hamiltonian operator includes besides the Ricci scalar an addition term proportional to the extrinsic curvature of the surface. In terms of the classical Gaussian and mean curvatures of the surface, $K$ and $M$, it reads
    \be
    H=-\frac{\hbar^2}{2m} g^{1/2} \partial_i (g^{ij}g^{1/2})\partial_j+
    \frac{\hbar^2}{m}\left(\lambda_1K+\lambda_2 M^2\right),
    \label{thin-layer}
    \ee
where $\lambda_1=-\lambda_2=1/2$. The study of physical systems described by the Hamiltonian operator~(\ref{thin-layer}) have been a focus of attention for decades \cite{Silva-2013,Encinosa-1998,Taira-2007,Atanasov-2009,Brandt-2015}. The generalizations of this Hamiltonian to particles interacting with electromagnetic fields, spin 1/2 particles, particles with position-dependent and anisotropic effective masses, and thin layers with small but finite thickness have been considered in \cite{Ferrari-2008,Atanasov-2011,Wang-2014, Souza-2018,Serafim-2019,Wang-2016}.

A more careful examination of the analysis leading to
Eq.~(\ref{thin-layer}) shows that in general the curvature
coefficients, $\lambda_1$ and $\lambda_2$, depend on the details of
the confining forces \cite{Kaplan-1997}, i.e., their values cannot
be determined from first principle. Viewing the system as a
second-class constrained system and employing Dirac's quantization
program for such systems leads to the same conclusion; the
Hamiltonian has the form (\ref{thin-layer}), but the curvature
coefficients cannot be uniquely determined from theoretical
considerations \cite{Golovnev-2009}. These observations provide
further motivation for the empirical determination of the curvature
coefficients by performing scattering experiments.

Ref.~\cite{pra-2018} provides a comprehensive treatment of the geometric scattering of a scalar particle moving on an asymptotically flat embedded surface. A possible candidate for realizing this system is a dilute electron gas formed on a bumpy surface. Motivated by the fact that such a condensed matter system would naturally involve defects, the authors of Ref.~\cite{ap-2019} explore the effects of point defects on the geometric scattering amplitude. This reveals the amplification of the geometric scattering effects by the point defects. The purpose of the present paper is to examine the influence of a set of parallel line defects on the geometric scattering amplitude for an embedded surface $S$ with cylindrical symmetry. In particular, we address the scattering problem for the Hamiltonian,
     \begin{equation}
    H=-\frac{\hbar^2}{2m} g^{1/2} \partial_i (g^{ij}g^{1/2})\partial_j+
    \frac{\hbar^2}{m}(\lambda_1K+\lambda_2 M^2)+ V_0,
    \label{H-gen-defect}
    \end{equation}
where $V_0$ is the potential,
    \be
    V_0(x,y)=\sum_{n=1}^N\xi_n\delta(x-a_n),
    \label{eq1}
    \ee
$(x,y)$ are the local Cartesian coordinates of the surface $S$, $N$ is the number of line defects, $\xi_n$ are real or complex coupling constants,  $\delta(x)$ denotes the Dirac delta function in one dimension, and $a_n$ are real numbers determining the position of the line defects.

For example consider the surface with the shape of a Gaussian bump \cite{Silva-2013} in the presence of a pair of parallel line defects. If the center of the bump lies in the region between the defects, they can serve as the walls of an effective resonator capable of producing multiple internal reflections of an incident wave. This suggests that the presence of the defects can produce a sizable amplification of the scattering of the wave due to the nontrivial geometry of the surface. The main purpose of the present article is to investigate the prospects of this amplification scheme for geometric scattering.

The organization of this article is as follows. In Sec.~2, we review the standard approach to potential scattering in two dimensions. In Sec.~3, we consider the scattering problem for the potential (\ref{eq1}) in a plane. In Sec.~4, we study the geometric scattering in the presence of line defects for the case that $S$ is an asymptotically flat surface with cylindrical symmetry. Here we consider the effects of the nontrivial geometry of the surface as a first-order perturbation of the case of lines defects in the plane. In Sec.~5, we confine our calculation to the surface of a Gaussian bump and provide a graphical demonstration of the behavior of the scattering cross section (length.) Finally, in Sec.~6, we present our concluding remarks.

\section{Potential scattering in two dimensions}

Consider the time-independent Schr\"odinger equation,
    \begin{equation}
    \label{sch-eq}
    H |\psi\kt=E |\psi\kt,
    \end{equation}
for a Hamiltonian operator of the Standard form, i.e.,
    \be
    H=\frac{\hat\bp^2}{2m}+V,
    \nn
    \ee
where $\hat\bp$ is the standard momentum operator in two dimensions, and $V$ is a scalar scattering potential. Scattering solutions $|\psi(\bk)\kt$ of (\ref{sch-eq}) satisfy the Lippmann-Schwinger equation,
    \begin{equation}
    \label{ls}
    |\psi(\bk)\kt=\ket{{\bk}}+
    G_0^+ (E) V|\psi(\bk)\kt,
    \end{equation}
where $\bk$ is the incident wavevector, and
    \begin{equation}
    \label{G-plus}
    G_0^+ (E):=
    \lim_{\epsilon\to 0^+}\frac{1}{E-\hat{\bp}^2/2m + i\epsilon}
    =\lim_{\epsilon\to 0^+}
    \int_{\R^2} d^2{\bk'} \frac{|\bk'\kt\br\bk'|}{E-\hbar^2k^{\prime 2}/2m+i\epsilon}.
    \end{equation}
We can express $\bk$ as $\bk=k_x\bfe_x+k_y\bfe_y=k(\cos\theta_0\bfe_x+
\sin\theta_0\bfe_y)$, where $\bfe_x$ and $\bfe_y$ are respectively the unit vectors along the $x$- and $y$-axes, $k:=\sqrt{2mE}/\hbar$ is the wavenumber, and $\theta_0$ is the incidence angle.

The integral kernel of $G_0^+ (E)$ in the position representation is the Green's function associated with the out-going solutions of (\ref{sch-eq}), i.e.,
    \[G^{+}_0(\bx,\bx'):=\br\bx|G_0^{+}(E)|\bx'\kt
    =-\frac{ im}{2\hbar^2} H_0^{(1)}(k|\bx-\bx'|),\]
where $H_0^{(1)}(x)$ is the zero-order Hankel function of the first kind.  Employing the well-known asymptotic expression for the latter, we can show that
    \be
    \br\bx|\psi(\bk)\kt\to
    \frac{1}{2\pi}\left[ e^{ i \textbf{k}\cdot \textbf{x} }+
    \ff({\bk}',{\bk})\,\frac{e^{i k r }}{\sqrt{r}}\right]
    \quad{\rm for} \quad r:=|\bx|\to\infty,
    \label{asymp}
    \ee
where $\bx=x\bfe_x+y\bfe_u$ marks the position of the detector, $(r,\theta)$ are polar coordinates of $\bx$, $\bk':=k\,\bx/r$, and $\ff({\bk}',{\bk})$ is the scattering amplitude for the potential $V$ which is given by
    \bea
    \ff({\bk}',{\bk})&:=&
    \frac{-2\pi m}{\hbar^2}\sqrt{\frac{2\pi i}{k}}\;
    \br\bk'|V|\psi(\bk)\kt.
    \label{f=def}
    \eea

Now, suppose that we can express $V$ as the sum of a scattering potential $V_0$ and a perturbation $\zeta V_1$;
    \be
    V=V_0+\zeta V_1,
    \label{pert-}
    \ee
where $\zeta$ is a real perturbation parameter. Then we can use the standard perturbation theory to obtain
the following Born series expansions for the scattering solution and scattering amplitude \cite{ap-2019}.
    \begin{align}
    \label{expand}
    &|\psi(\bk)\kt=\sum_{n=0}^{\infty} \zeta^n |\psi_n(\bk)\kt,
    &&\ff(\bk',\bk)=\sum_{n=0}^\infty \,\zeta^n \ff_n(\bk',\bk),
    |\psi_n(\bk)\kt,
    \end{align}
where
    \bea
    \label{unpertured wave}
    |\psi_n(\bk)\kt&:=&\left\{\begin{array}{ccc}
    \left[1-G_0^+(E) V_0 \right]^{-1} |\bk\kt &{\rm for}& n=0,\\[6pt]
    \left[1-G_0^+(E) V_0 \right]^{-1} G_0^+(E)
    V_1 |\psi_{n-1}(\bk)\kt &{\rm for}& n\geq 1,
    \end{array}\right.\\[6pt]
    \ff_n(\bk',\bk)&:=&\frac{-2\pi m}{\hbar^2}\sqrt{\frac{2\pi i}{k}}\times
    \left\{\begin{array}{ccc}
    \br\bk'|V_0|\psi_0(\bk)\kt & {\rm for} & n=0,\\[6pt]
    \br\bk'|V_0|\psi_n(\bk)\kt+\br\bk'|V_1|\psi_{n-1}(\bk)\kt
    & {\rm for} & n\geq 1.\end{array}\right.
    \label{fn-def}
    \eea

The first Born approximation corresponds to neglecting all but the first two terms in the series (\ref{expand}). This gives $\ff(\bk',\bk)\approx \ff_0(\bk',\bk)+\zeta\, \ff_1(\bk',\bk)$, where
    \bea
    \ff_0(\bk',\bk)&:=&\frac{-2\pi m}{\hbar^2}\sqrt{\frac{2\pi i}{k}}\br\bk'|V_0|\psi_0(\bk)\kt,
    \label{f-zero}\\
    \ff_1(\bk',\bk)&:=&\frac{-2\pi m}{\hbar^2}\sqrt{\frac{2\pi i}{k}}
    \left[\br\bk'|V_0|\psi_1(\bk)\kt+\br\bk'|V_1|\psi_{0}(\bk)\kt\right].
    \label{f-one}
    \eea
Note that $|\psi_0(\bk)\kt$ and $\ff_0(\bk',\bk)$ are respectively
an exact scattering solution of the time-independent Schr\"odinger
equation and the corresponding exact scattering amplitude for the
unperturbed potential $V_0$.

Setting $n=1$ in (\ref{unpertured wave}) and using the result together with the geometric series expansion for $[1-G^+_0(E)V_0]^{-1}$ and $[1-V_0G^+_0(E)]^{-1}$, we can show that
    \bea
    \br\bk'|V_0|\psi_1(\bk)\kt+\br\bk'|V_1|\psi_0(\bk)\kt&=&
    \br\bk'|\left\{V_0[1-G^+_0(E)V_0]^{-1}G^+_0(E)+1\right\}V_1|\psi_0(\bk)\kt\nn\\
    &=& \br\bk'|[1-V_0G^+_0(E)]^{-1}V_1|\psi_0(\bk)\kt\nn\\
    &=&\br\tilde\psi_0(\bk')|V_1|\psi_0(\bk)\kt,
    \label{VV=tilde}
    \eea
where
    \begin{align}
    &|\tilde\psi_0(\bk)\kt:=[1-G^-_0(E)V_0^\dagger]^{-1}|\bk\kt,
    \label{psi-tilde}
    \end{align}
and $G^-_0(E):=G^+_0(E)^\dagger$. Clearly,
    \be
    \br\bx|G^-_0(E)|\bx'\kt=\br\bx'|G^+_0(E)|\bx\kt^*=
    \frac{ im}{2\hbar^2} {H_0^{(1)}(k|\bx-\bx'|)}^{\!*}=\br\bx|G^+_0(E)|\bx'\kt^*.
    \label{G-minus}
    \ee
Substituting (\ref{VV=tilde}) in (\ref{f-one}) yields
    \be
    \ff_1(\bk',\bk):=\frac{-2\pi m}{\hbar^2}\sqrt{\frac{2\pi i}{k}}
    \br\tilde\psi_0(\bk')|V_1|\psi_0(\bk)\kt.
    \label{f-one-2}
    \ee

For the scattering problem we consider, $V_0$ is the potential (\ref{eq1}), which models the line defects, and the perturbation takes the form
    \be
    \zeta V_1:=H-H_0,
    \ee
where, in the position representation, $H$ and $H_0$ are respectively given by (\ref{H-gen-defect}) and
    \be
    H_0:=-\frac{\hbar^2}{2m}\nabla^2+V_0.
    \label{H0}
    \ee

\section{Scattering by parallel line defects in a plane}

Consider the case that $S$ is the Euclidean plane. Then the Hamiltonian (\ref{H-gen-defect}) reduces to (\ref{H0}), and in the position representation the time-independent Schr\"odinger equation reads
    \begin{equation}
    \Big[-\partial_x^2-\partial_y^2+\sum_n^N \fz_n \delta (x-a_n)\Big]\psi_0(x,y)=k^2\psi_0(x,y)\,,
    \label{eq2}
    \end{equation}
where $\fz_n:=2m\xi_n/\hbar^2$. The solution of the scattering problem defined by (\ref{eq2}) is equivalent to finding the scattering amplitude, $\ff_0(\bk',\bk)$, for the unperturbed potential $V_0$.

Because the potential term in (\ref{eq2}) does not depend on $y$, we can easily solve this equation by separation of variables. In particular, introducing $\phi(y):=e^{ik_y y}/2\pi$ and demanding that
    \be
    \br\bx|\psi_0(\bk)\kt=\br x,y|\psi_0(\bk)\kt=\chi(x)\phi(y),
    \label{eq101}
    \ee
satisfies the Schr\"odinger equation (\ref{eq2}) for some auxilliary function $\chi$, we find
    \be
    \chi''(x)+k_x^2\chi(x)=\sum_n^N \fz_n \delta (x-a_n)\chi(a_n).
    \label{eq102}
    \ee
We need to find a solution of this equation such that   $\br\bx|\psi_0(\bk)\kt$, as given by (\ref{eq101}), solves the Lippmann-Schwinger equation,
    \be
    \br\bx|\psi_0(\bk)\kt=\br\bx|\bk\kt+\br\bx|G^+_0(E)V_0|\psi_0(\bk)\kt.
    \label{L-Sch-eq}
    \ee

To simplify the second term on the right-hand side of (\ref{L-Sch-eq}), we identify the Hilbert space $L^2(\R^2)$ of square-integrable functions of $\bx=(x,y)$ with $\sH_1\otimes\sH_2$, where $\sH_1$ and $\sH_2$ are respectively the Hilbert space of the square-integrable functions of $x$ and $y$. This allows us to express $|\psi_0(\bk)\kt$ and $V_0$ in the form,
    \begin{align}
    &|\psi_0(\bk)\kt = |\chi, \phi\kt:=|\chi\kt \otimes | \phi\kt,
    \label{psi-zero}\\
    &V_0=\sum_{n=1}^N \xi_n  |a_n\kt\br a_n| \otimes {I}_2,
    \label{v0-n}
    \end{align}
where ${I}_2$ is the identity operator for $\sH_2$. We can use these equations together with (\ref{G-plus}), and $\phi(y):=\br y|k_y\kt/\sqrt{2\pi}$ to show that
    \bea
    \br x,y|G_0^+(E)V_0|\psi_0(\bk)\kt&=&
    \frac{1}{2\pi}\sum_{n=1}^N\xi_n \chi(a_n) \int_{-\infty}^\infty d\tilde k_x e^{-ia_n\tilde k_x}
    \br x,y|G^+_0(E)|\tilde k_x,k_y\kt \nn\\
    &=&\phi(y)\sum_{n=1}^N\fz_n \chi(a_n)\sG(x-a_n),
    \label{eq-391}
    \eea
where $\sG$ is the Green's function for the operator $\partial_x^2+k_x^2$ that is given by
    \begin{align}
    \sG(x-x'):= \lim_{\epsilon\to 0^+}\br x|\left(-\hat k_x^2+k_x^2+i\epsilon\right)^{-1}|x'\kt=
    -\frac{i e^{i k_x|x-x'|}}{2k_x},
    \label{eq105}
    \end{align}
$\hat k_x:=\hat p_x/\hbar$, and $\hat p_x$ is the $x$-component of the momentum operator $\hat\bp$.

Next, we substitute (\ref{eq-391}) in (\ref{L-Sch-eq}) and use
(\ref{eq105}) to show that
    \begin{equation}
    \chi(x)= e^{ik_x x}-\frac{i  }{2k_x}\sum_n^N \fz_n e^{ik_x|x-a_n|}\chi(a_n).
    \label{eq7}
    \end{equation}
Setting  $x=a_m$, with $m=1,\dots,N$, in this equation, we arrive at the following system of linear equations for $\chi(a_n)$.
    \begin{equation}
    \sum_{n=1}^N T_{nm} \chi(a_n)=e^{ik_x a_m},
    \label{eq9}
    \end{equation}
where
    \begin{align}
    T_{nm}:=\delta_{nm}+
    \frac{i \fz_n}{2k_x} e^{ik_x|a_m-a_n|}=
    \left\{\begin{array}{ccc}
    1+ \frac{i  \fz_n}{2k_x}   &\text{for} & n=m,\\[6pt]
    \frac{i \fz_n }{2k_x}\,  e^{ik_x|a_m-a_n|} &\text{for}& n\neq m,
    \end{array}\right.
    \label{eq10}
    \end{align}
and $\delta_{nm}$ is the Kronecker delta symbol.  According to (\ref{eq7}) and (\ref{eq9}), we can express the scattering solution (\ref{eq101}) of the Schr\"odinger equation (\ref{eq2}) in the form,
    \be
    \br\bx|\psi_0(\bk)\kt=\frac{1}{2\pi}\left[ e^{i\bk\cdot\bx}-i
    \sum_{m,n=1}^N e^{i k_x a_m}A^{-1}_{mn}\,
    e^{i(k_x |x-a_n|+k_y y)}\right],
    \label{eq12}
    \ee
where  $A^{-1}_{mn}$ are the entries of the inverse of the matrix $\bA:=[A_{mn}]$ with
    \be
    A_{mn}:=\frac{2 k_x T_{mn}}{\fz_m}=
    \frac{2 k_x\,\delta_{mn}}{\fz_m}+
    i e^{ik_x|a_m-a_n|}=
    \left\{\begin{array}{ccc}
    \frac{2k_x}{\fz_m}+ i  &\text{for} & n=m,\\[6pt]
    i  e^{ik_x|a_m-a_n|} &\text{for}& n\neq m.
    \end{array}\right.
    \label{eq122}
    \ee

In order to determine the scattering amplitude $\ff_0(\bk',\bk)$, we should derive the asymptotic expression for the right-hand side of (\ref{eq12}) and put it in the form (\ref{asymp}).  We present the details of this calculation in Appendix~A. Its final result is:
    \bea
    \ff_0(\bk',\bk)&=&
    -\sqrt{\frac{2\pi i}{k }}\sum_{n,m=1}^N A^{-1}_{mn}
    \Big[e^{i k_x(a_m-a_n)} \delta(\theta-\theta_0)+
    e^{i k_x(a_m+a_n)} \delta(\theta+\theta_0-\pi)\Big],
    \label{eq20}\\
    &=&\sqrt{\frac{2\pi}{k}}\,e^{-i\pi/4}
    \Big[\ft^+(\bk)\delta(\theta-\theta_0)+\ft^-(\bk)\delta(\theta+\theta_0-\pi)\Big],
    \label{eq20n}
    \eea
where we have introduced
    \begin{align}
    &\ft^+(\bk):=-i\sum_{n,m=1}^N A^{-1}_{mn}e^{i k_x(a_m-a_n)}=
    -i\sum_{n,m=1}^N A^{-1}_{mn}\cos[k_x(a_m-a_n)],\\
    &\ft^-(\bk):=-i\sum_{n,m=1}^N A^{-1}_{mn}e^{i k_x(a_m+a_n)},
    \end{align}
and used the fact that $\bA^{-1}$ is a symmetric matrix. According to (\ref{eq20n}), the scattered wave consists of a transmitted part that travels along the same direction as the incident wave ($\theta=\theta_0$) and a reflected part that returns to $x=-\infty$ along a ray with inclination $\theta=\pi-\theta_0$.

\section{Geometric scattering for a surface with line defects}

To determine the geometric scattering properties of our system, we express the Hamiltonian operator (\ref{H-gen-defect}) as the sum of the geometric and non-geometric contributions,
    \be
    H=H_0+\zeta\,V_1,
    \label{eqn-22}
    \ee
where $H_0$ is given by (\ref{H0}), $\zeta\,V_1:=H-H_0$, and $\zeta$ is an arbitrary real parameter that we have introduced to keep track of the strength of the geometric contributions. In view of (\ref{H-gen-defect}) and (\ref{H0}),
    \begin{equation}
    \zeta\br\bx'|V_1|\bx\kt=\frac{\hbar^2}{2m}\,\cL_{\bx'}\,\delta(\bx'-\bx),
    \label{eq23}
    \end{equation}
where $\cL_{\bx}$ is the differential operator,
    \begin{equation}
    \cL_{\bx}:=\left[g_0^{ij}(x)-g^{ij}(x)\right]\partial_i \partial_j
    -\frac{\partial_i[\sqrt{g(x)} g^{ij}(x)]}{\sqrt{g(x)}}\partial_j 
    + 2\lambda_1 K(x)+2 \lambda_2 M(x)^2,
    \label{eq24}
    \end{equation}
and $g_0^{ij}$ are  the components of the inverse of the Euclidean metric tensor $\fg_0$.\footnote{Whenever $(x^1,x^2)$ are Cartesian coordinates, $g_0^{ij}=\delta_{ij}$.}

In what follows, we identify the geometric contributions represented by $\zeta\,V_1$ as a perturbation and use first-order perturbation theory to account for its scattering effects. In particular, we express the scattering amplitude as
    \begin{equation}
    \ff(\bk',\bk)\approx \ff_0(\bk',\bk)+\zeta\, \ff_1(\bk',\bk),\nn
    \end{equation}
where $\approx$ means that we neglect quadratic and higher order terms in powers of $\zeta$, $\ff_0(\bk',\bk)$ is given by \eqref{eq20}, and $\zeta\, \ff_1(\bk',\bk)$ represents the first order contributions. The latter quantifies the geometric scattering effects. Because $\ff_0(\bk',\bk)$ vanishes for angles $\theta$ other than $\theta_0$ and $\pi-\theta_0$,
    \[ \ff(\bk',\bk)\approx \zeta\, \ff_1(\bk',\bk)~~~{\rm for}~~~\theta\notin\{\theta_0,\pi-\theta_0\}.\]
This implies that the scattering of the wave along generic directions is essentially geometric in nature. This is in contrast with the scattering by a surface with point defects \cite{ap-2019}.

In view of \eqref{f-one-2} and (\ref{eq23}),
    \be
    \zeta\,\ff_1(\bk',\bk)=-\pi\sqrt{\frac{2\pi i}{k}}
    \int_{\R^2}d^2\bx'\br\bx'|\tilde\psi_0(\bk')\kt^*\cL_{\bx'}\br\bx'|\psi_0(\bk)\kt.
    \label{eq26}
    \ee
Therefore, in order to determine $\ff_1(\bk',\bk)$, we need to compute $\br\bx|\tilde\psi_0(\bk)\kt$.
We can use (\ref{psi-tilde}) to identify the latter with the solution of the Lippmann-Schwinger equation,
    \[\br\bx|\tilde\psi(\bk)\kt=\br\bx|\bk\kt+\br\bx|G_0^-(E)V_0^\dagger|\tilde\psi(\bk)\kt.\]
In view of the analogy between this equation and (\ref{L-Sch-eq}), we can use the analysis leading to the expression (\ref{eq12}) for $\br\bx|\psi_0(\bk)\kt$ together with Eq.~(\ref{G-minus}) to show that
    \be
    \br\bx|\tilde\psi_0(\bk)\kt=\frac{1}{2\pi}\left[ e^{i\bk\cdot\bx}+i
    \sum_{m,n=1}^N e^{i k_x a_m}A^{-1*}_{mn}\,
    e^{i(-k_x |x-a_n|+k_y y)}\right].
    \label{eq12-tilde}
    \ee

Next, we substitute \eqref{eq12} and (\ref{eq12-tilde}) in (\ref{eq26}) to obtain
    \begin{equation}
    \label{eq30}
    \zeta \ff_1(\bk',\bk)=-\frac{1}{2}\sqrt{\frac{i}{2\pi k}}\left[I_{0}
    -i\!\sum_{m,n=1}^N\!\!\left(A^{'-1}_{mn}I_{mn}+A^{-1}_{mn} J_{mn}\right)
    -\!\!\!\sum_{m,n,m',n'=1}^N \!\!\!A^{'-1}_{mm'}A^{-1}_{nn'}I_{m m' n n'}\right],
    \end{equation}
    where $A^{'-1}_{mn}$ stands for $A^{-1}_{mn}$ with $k_x$ replaced with $k_x'$, and $I_0$, $I_{mn}$, $J_{mn}$, and $I_{mm'nn'}$ are complex coefficients given by
    \bea
    \label{eq31}
    I_{0}&:=&\int\limits_{\R^2}d^2\bx'\:e^{-i\bk'\cdot\bx'}\cL_{\bx'}\, e^{i\bk\cdot\bx'},\\
    \label{eq32}
    I_{mn}&:=&\int\limits_{\R^2}d^2\bx'
    \Big(e^{-i k'_{x'} a_m}\ e^{-ik'_{y'} {y'}}e^{-ik'_{x'}|{x'}-a_n|}\Big)\,\cL_{\bx'}\: e^{i\bk\cdot\bx'},\\
    \label{eq33}
    J_{mn}&:=&\int\limits_{\R^2}d^2\bx'
    e^{-i\bk'\cdot\bx'}\cL_{\bx'} \Big(e^{i k_{x'} a_m}\ e^{ik_{y'} {y'}}e^{ik_{x'}|{x'}-a_n|}\Big),\\
    \label{eq34}
    I_{mm'nn'}&:=&\int\limits_{\R^2}d^2\bx'\,
    \Big(e^{-i k'_{x'} a_{m'}}\ e^{-ik'_{y'} {y'}}e^{-ik'_{x'}|{x'}-a_{m}|}\Big) \cL_{\bx'}\Big(e^{i k_{x'} a_{n'}}\ e^{ik_{y'} {y'}}e^{ik_{x'}|{x'}-a_n|}\Big).
    \eea

For a general embedded surface $S$, obtaining useful explicit expressions for $I_{0}, I_{mn}, J_{mn}$, and $I_{mm'nn'}$  turns out to be intractable. For the reason, in the remainder of this article we confine our attention to the cases where $S$ has cylindrical symmetry. More precisely, we let $(r,\theta,z)$ to label the cylindrical coordinates in $\R^3$, and suppose that $S$ is the subset of $\R^3$ determined by
    \begin{equation}
    z=f(r),
    \end{equation}
where $f:[0,\infty)\to\R$ is a smooth function satisfying
    \begin{equation}
    \lim_{r \to \infty} \dot{f}(r)=\lim_{r \to 0} \dot{f}(r)=0,
    \label{eq36}
    \end{equation}
and an overdot stands for a derivative with respect to $r$, \cite{pra-2018}.

We can identify $(r,\theta)$ with the polar coordinates in $\R^2$ and use them as local coordinates on $S$, so that $x^1=r$ and $x^2=\theta$. In these coordinates the components of the metric tensor take the form \cite{pra-1996}:
    \begin{align}
    &g_{11}=1+\dot{f}^2, && g_{12}=g_{21}=0, &&g_{22}=r^2,
    \label{metric=}
    \end{align}
and we can respectively express the Gaussian and mean curvatures of $S$ as
    \begin{equation}
    K=\frac{G\dot{G}}{r}, \qquad\qquad\qquad M=\frac{1}{2}\left(\frac{G}{r}+\dot{G}\right),
    \label{curvatures}
    \end{equation}
 where
    \begin{equation}
    G:=\frac{\dot{f}}{\sqrt{1+\dot{f}^2}}.
    \label{G-def}
    \end{equation}
According to (\ref{curvatures}) and (\ref{G-def}), $K$ and $M$ are regular (non-singular) functions of $r$ provided that $f'(r)/r$ and $f''(r)$ tend to finite limits as $r \rightarrow 0$, \cite{pra-2018}.

Next, we employ (\ref{metric=}) and (\ref{curvatures}) to compute the differential operator (\ref{eq24}). This gives
    \begin{equation}
    \begin{array}{lll}
    \cL_\textbf{x}= G^2\left[\partial^2_{r}+\frac{1}{r}\left(1+\frac{r\dot{G}}{G}\right)\partial_{r}+
    \frac{2\lambda_1\dot{G}}{r\, G} +\frac{\lambda_2}{2r^2} \left(1+
    \frac{r\dot{G}}{G}\right)^{\!2}\right].
    \end{array}
    \label{eq40}
    \end{equation}
The use of this relation for the purpose of computing the coefficients $I_{0}, I_{mn}, J_{mn}$, and $I_{mm'nn'}$  that appear in the expression (\ref{eq30}) for the geometric scattering amplitude encounters major technical difficulties. To circumvent these we restrict to the case where $S$ is a Gaussian bump. In particular, we set
    \begin{equation}
    \label{eq47}
    f(r)=\delta\, e^{-r^2/2\sigma^2},
    \end{equation}
where $\delta$ and $\sigma$ are real parameters with the dimension of length that respectively represent the height and width of Gaussian bump, and demand that $(\delta/\sigma)^2\ll 1$. The latter allows us to expand the terms contributing to the integrals in (\ref{eq31}) -- (\ref{eq34}) in powers of
    \[\eta:=\frac{\delta^2}{\sigma^2}\]
and ignore the quadratic and higher order terms. To evaluate these integrals we choose a coordinate system in which $\Delta \boldsymbol{k} =\boldsymbol{k}-\boldsymbol{k}'$ lies along the $x'$-axis.

If we use $\Theta$ (respectively $\theta$) to denote the angle between $\boldsymbol{k}$ and  $\boldsymbol{k}'$ (respectively $\bk'$ and the $x'$-axis), we can show that $\theta=(\pi+\Theta)/2$ and $|\bk'-\bk|=2ks$, where
    \begin{equation}
    s:=\sin(\Theta/2).
    \nn
    \end{equation}
Making use of these relations and Eq.~(\ref{eq31}), we find \cite{pra-2018,ap-2019}:
    \begin{equation}
    \label{eq51}
    I_0=\frac{\pi\,\eta\, e^{-s^2\fK^2}}{2} \Big[
    (4\lambda_1s^2-1)\fK^2+\lambda_2(s^4\fK^4+2)\Big]+\cO(\eta^2).
    \end{equation}
where $\fK:=k\sigma$, and $\cO(\eta^d)$ stands for terms of order $d$ and higher in powers of $\eta$.

The evaluation of the integrals in (\ref{eq32}) -- (\ref{eq34}) poses another difficulty, namely that their integrands involve functions of Cartesian coordinates $(x',y')$. We therefore perform a coordinate transformation to express the right-hand side of (\ref{eq40}) in Cartesian coordinates.  Inserting the result in (\ref{eq32}) -- (\ref{eq34}) and using various properties of Bessel functions and the identities,
    \begin{equation}
    \frac{d|x|}{dx} =\text{sgn}(x), \qquad\qquad
    \frac{d^2|x|}{dx^2} =2\delta(x),\nn
    \end{equation}
we can express $I_{mn}, J_{mn}$, and $I_{mm'nn'}$ in terms of the error and complementary error functions. We give the resulting expressions in Appendix~B. Substituting these in \eqref{eq30}, we obtain the scattering amplitude for the Gaussian bump \eqref{eq47} in the presence of $N$ parallel line defects located at $x=a_n$ with $n=1,..,N$.

Because of the complicated structure of the analytic formula for the scattering amplitude, we explore its implications graphically. For this purpose we imagine that our two-dimensional scattering system is realized in a dilute electron gas maintained on a Gaussian bump (\ref{eq47}) with identical line defects located at $x=a_n$. We approximate the delta function potential $\xi_n\delta(x-a_n)$ modeling the defects with the barrier potential,
    \be
    V_n(x,y):=\left\{\begin{array}{ccc}
    V_0 & {\rm for} & |x-a_n|\leq \rho/2,\\
    0 & {\rm for} & |x-a_n|> \rho/2,
    \end{array}\right.
    \label{barrier-pot}
    \ee
where $V_0:=\xi_n/\rho$ and $\rho$ are respectively the height and width of the barrier. For this approximation to be reliable, $V_0$ must be much larger than the energy $E:=(\hbar k)^2/2m$ of the incident electron, and $\rho$ must be much smaller than the length scales of the problem (the de~Broglie wavelength $\lambda:=2\pi/k=2\pi\hbar/\sqrt{2mE}$ and the width of the Gaussian bump $\sigma$), i.e.,
    \begin{align}
    &V_0\gg E,&&\rho\ll\lambda, && \rho\ll\sigma.
    \label{condi}
    \end{align}

For the geometric scattering effects to be significant, we should consider the scattering of the incident waves with wavelengths $\lambda$ that are of the same order of magnitude as $\sigma$. This means that $\fK:=k\sigma$ is of the order of 1. For these waves, we only need to satisfy the first two of the conditions listed in (\ref{condi}). We can express the first of these condition as $\fz_n=2m\xi_n/\hbar^2\gg k^2\rho$. Therefore it will be fulfilled, if $\sigma\fz_n\gg k\rho$. Note also that the second condition in (\ref{condi}) is equivalent to $k\rho\ll 2\pi$.

In the following, we set
    \begin{align}
    &V_0\approx 1~{\rm eV}, &&\rho\approx 1~{\rm nm}, && \fz_n=\sigma^{-1},
    \label{spec}
    \end{align}
and suppose that the effective mass of the electron is given by $m\approx10^{-2}m_{\rm e}$. Then it is easy to show that $E\ll V_0$ will imply $k\rho\ll 1$. For example, for $E\approx10^{-3}~{\rm eV}$ we find $k\rho\approx 0.02$.

\begin{figure}[ht]
    \begin{center}
        \includegraphics[scale=.43]{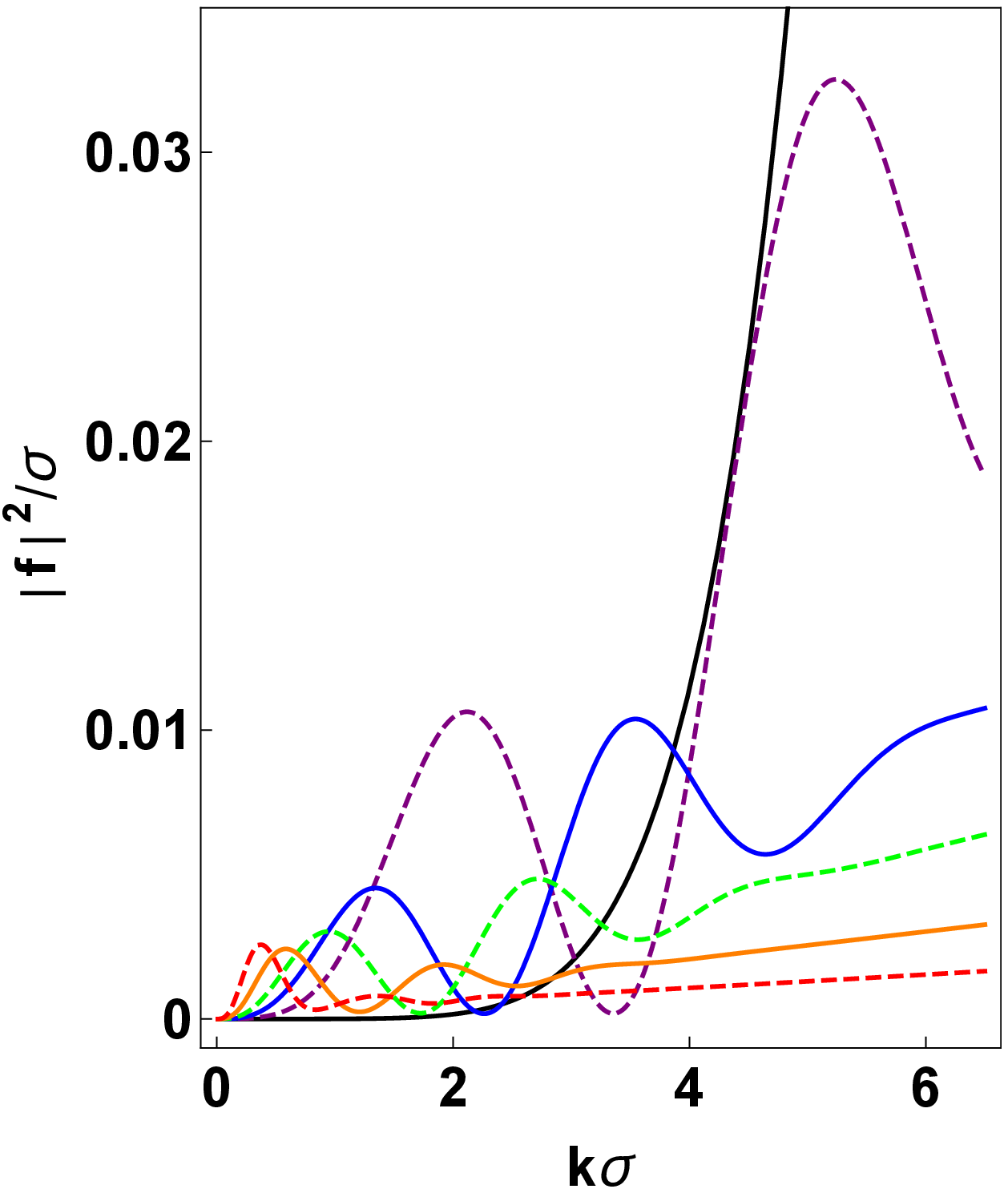}~~~
        \includegraphics[scale=.43]{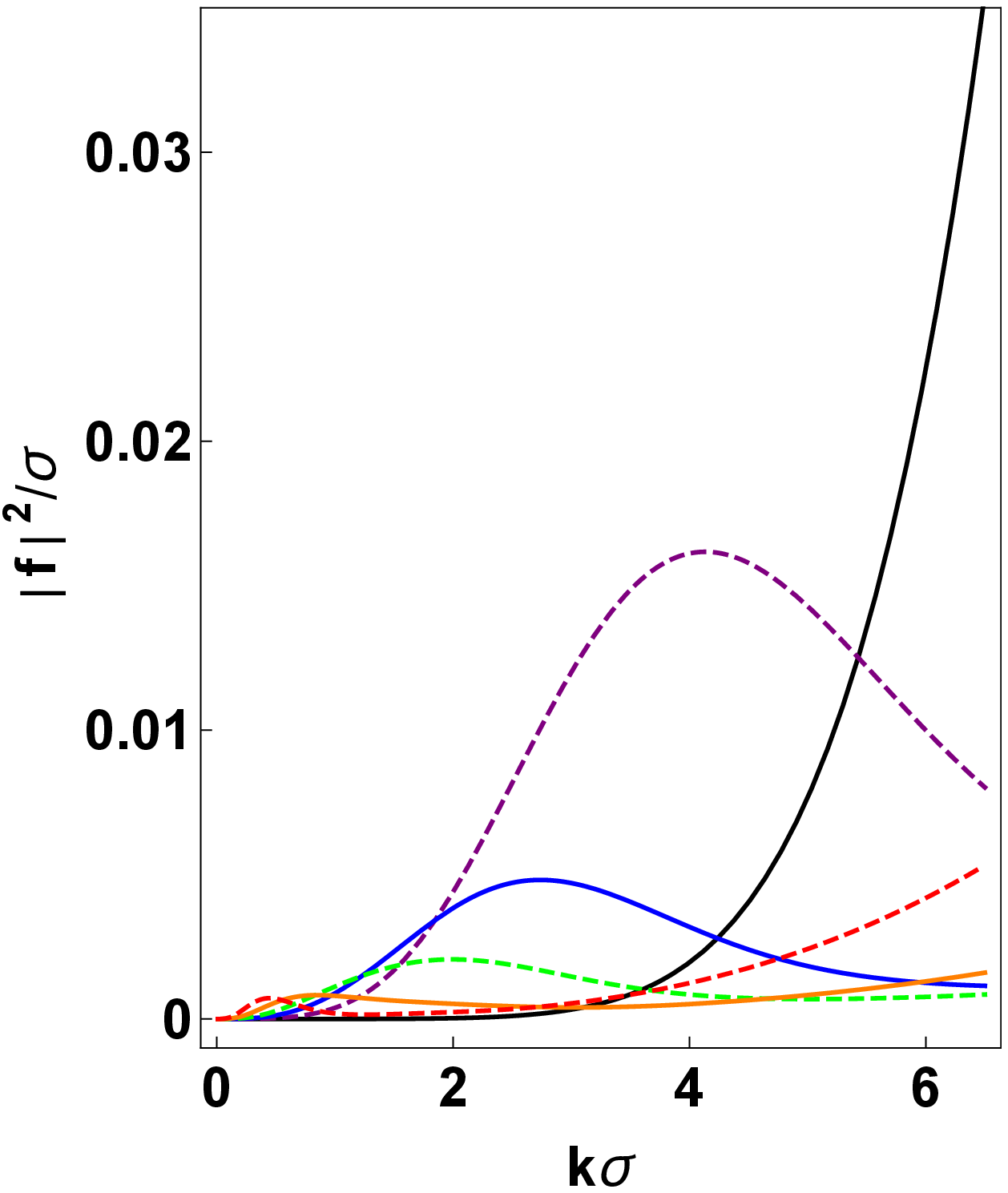}~~~
        \includegraphics[scale=.43]{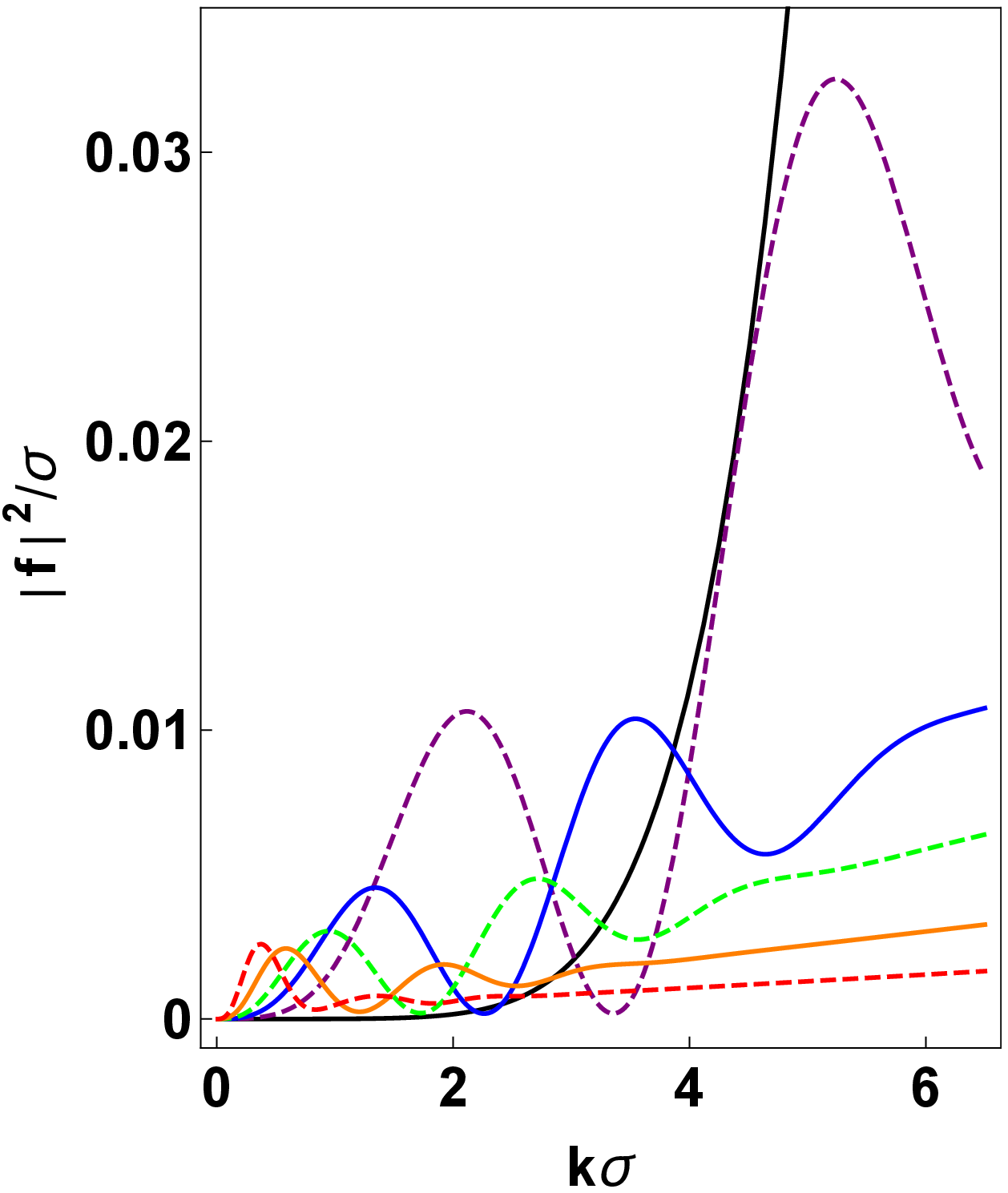}
        \caption{Plots of $|f(\boldsymbol{k}',\boldsymbol{k})|^2/\sigma$  as functions of $k\sigma$ for the Gaussian bump \eqref{eq47} with a line defect at $x=-3\sigma$ (on the left),  $x=0$ (in the middle), and $x=3\sigma$ (on the right) for $\theta_0=0^\circ$, $\eta=0.1$, $\sigma\fz_1=1$, $\lambda_1=-\lambda_2=1/2$, and different values of $\theta$, namely $\theta=5^\circ$ (black), $\theta=30^\circ$ (dashed purple), $45^\circ$ (blue), $60^\circ$ (dashed green), $90^\circ$ (orange), and $175^\circ$ (dashed red).}
        \label{fig1}
    \end{center}
    \end{figure}

\begin{figure}[ht]
    \begin{center}
        \includegraphics[scale=.43]{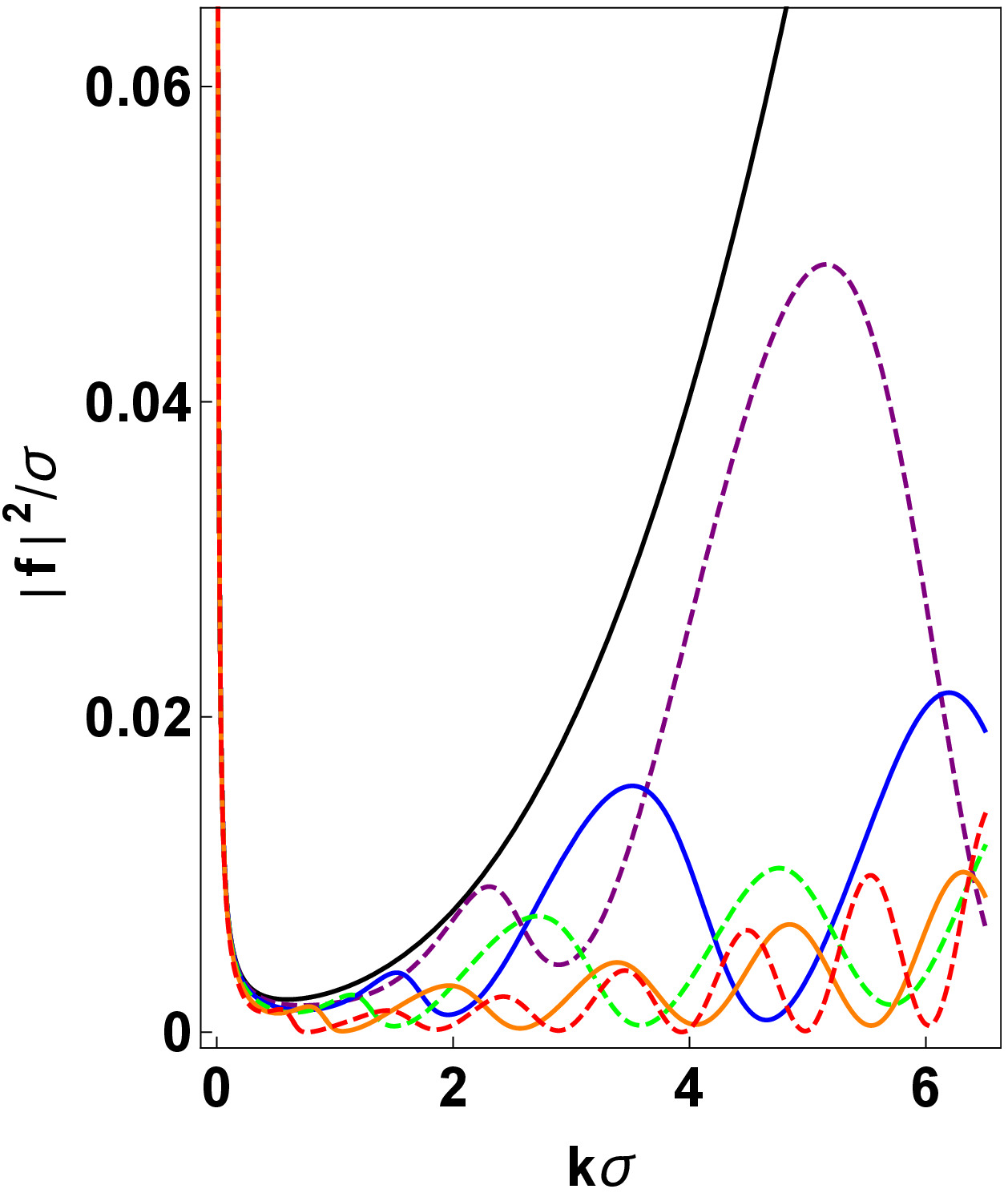}
        \includegraphics[scale=.43]{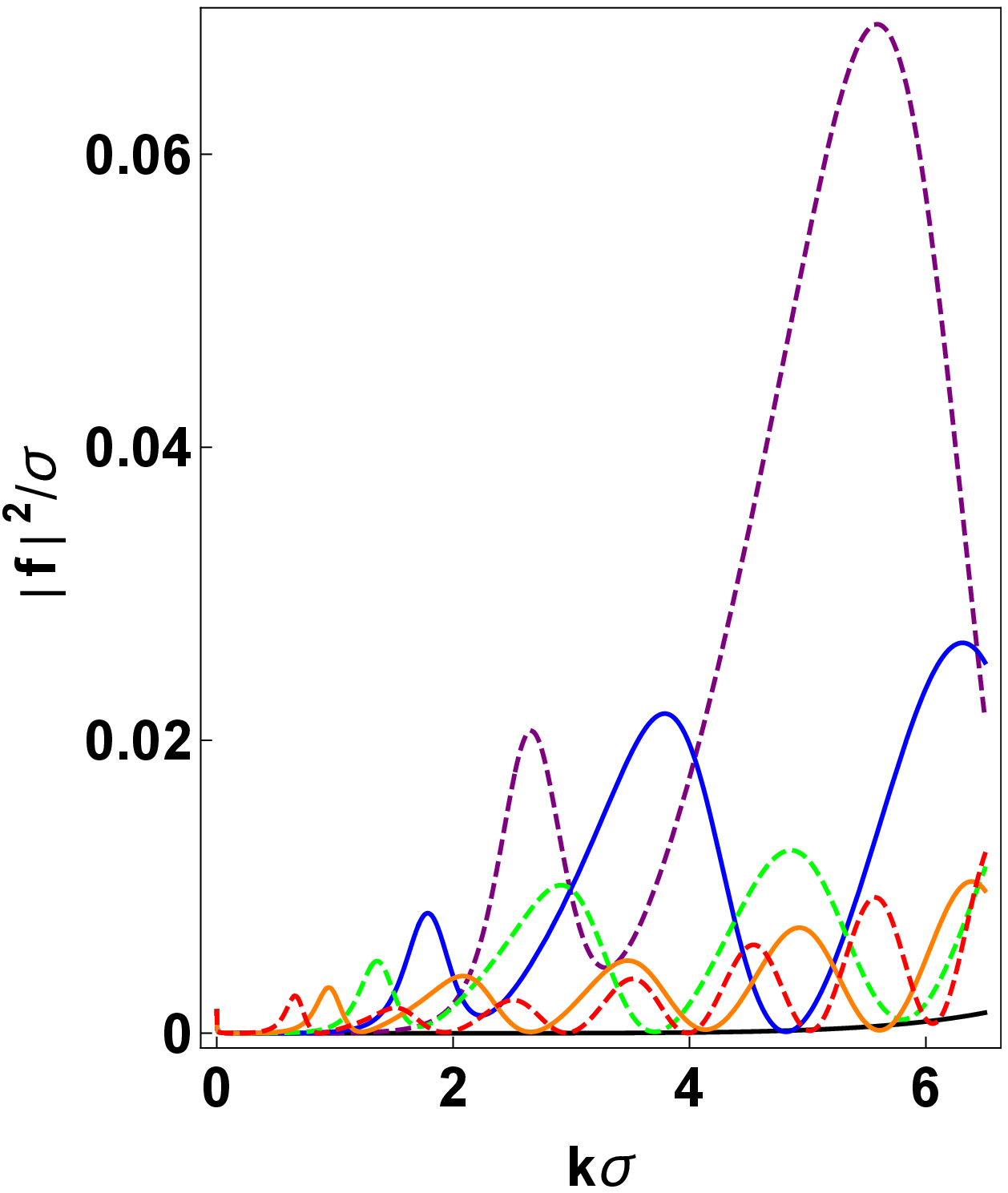}
        \includegraphics[scale=.43]{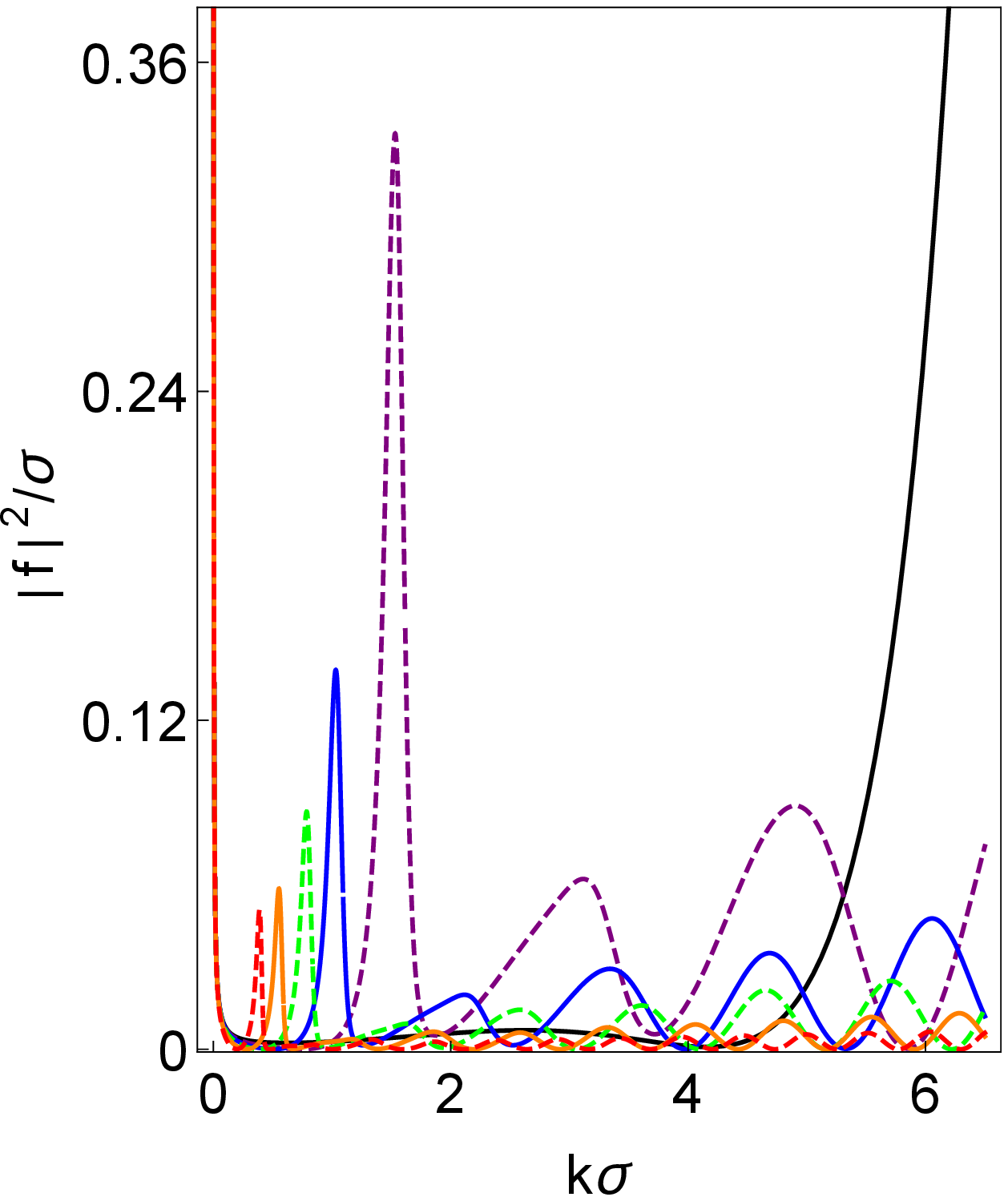}
        \caption{Plots of $|f(\boldsymbol{k}',\boldsymbol{k})|^2/\sigma$  as functions of $k\sigma$ for the Gaussian bump \eqref{eq47} with two line defects at $x=-3\sigma$ and $x=0$ (on the left),  $x=0$ and $x=3\sigma$ (in the middle), and $x=\pm 3\sigma$ (on the right) for $\theta_0=0^\circ$, $\eta=0.1$, $\sigma\fz_1=1$, $\lambda_1=-\lambda_2=1/2$, and different values of $\theta$, namely $\theta=5^\circ$ (black), $\theta=30^\circ$ (dashed purple), $45^\circ$ (blue), $60^\circ$ (dashed green), $90^\circ$ (orange), and $175^\circ$ (dashed red).}
        \label{fig2}
    \end{center}
    \end{figure}

Figures~\ref{fig1} and \ref{fig2} show the plots of the differential cross section $|\ff(\bk',\bk)|^2$ as a function of $\fK=k\sigma$ for a Gaussian bump in the presence of one or two line defects at different scattering angles $\theta$. Here we have taken $\lambda_1=-\lambda_2=1/2$, which is the prescription provided by the thin-layer quantization scheme \cite{daCosta-1981}. According to Figure~\ref{fig1}, the geometric scattering effects are more pronounced when the line defect does not pass through the center of the bump. Furthermore, the geometric scattering cross section corresponding to a line defect placed to the left of the bump is almost identical to that of a line defect placed to its right. This seems to suggest that the differential cross section is invariant under a reflection with respect to the $y$-axis. Numerical evidence turns out not to support this assertion; such a reflection produces a minute change in the cross section which is too small to be visible in our plots.
Figure~\ref{fig2} shows that the geometric scattering cross section
takes much larger values when the line defects are symmetrically
positioned with respect to the bump. This confirms our expectation
that a pair of parallel line defects can function as a resonator
capable of amplifying geometric scattering effects.

\begin{figure}
    \begin{center}
        \includegraphics[scale=.43]{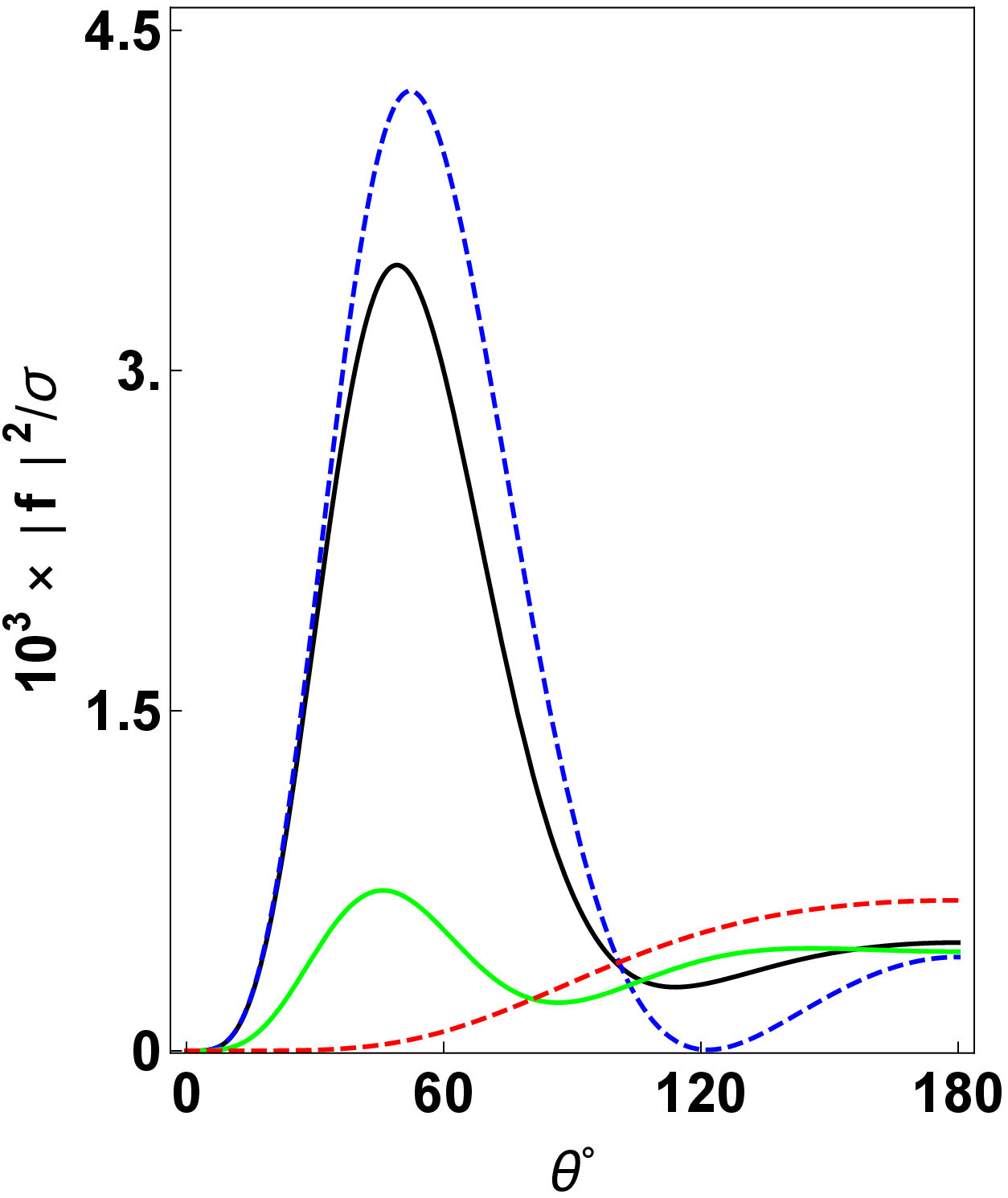}~~~
        \includegraphics[scale=.43]{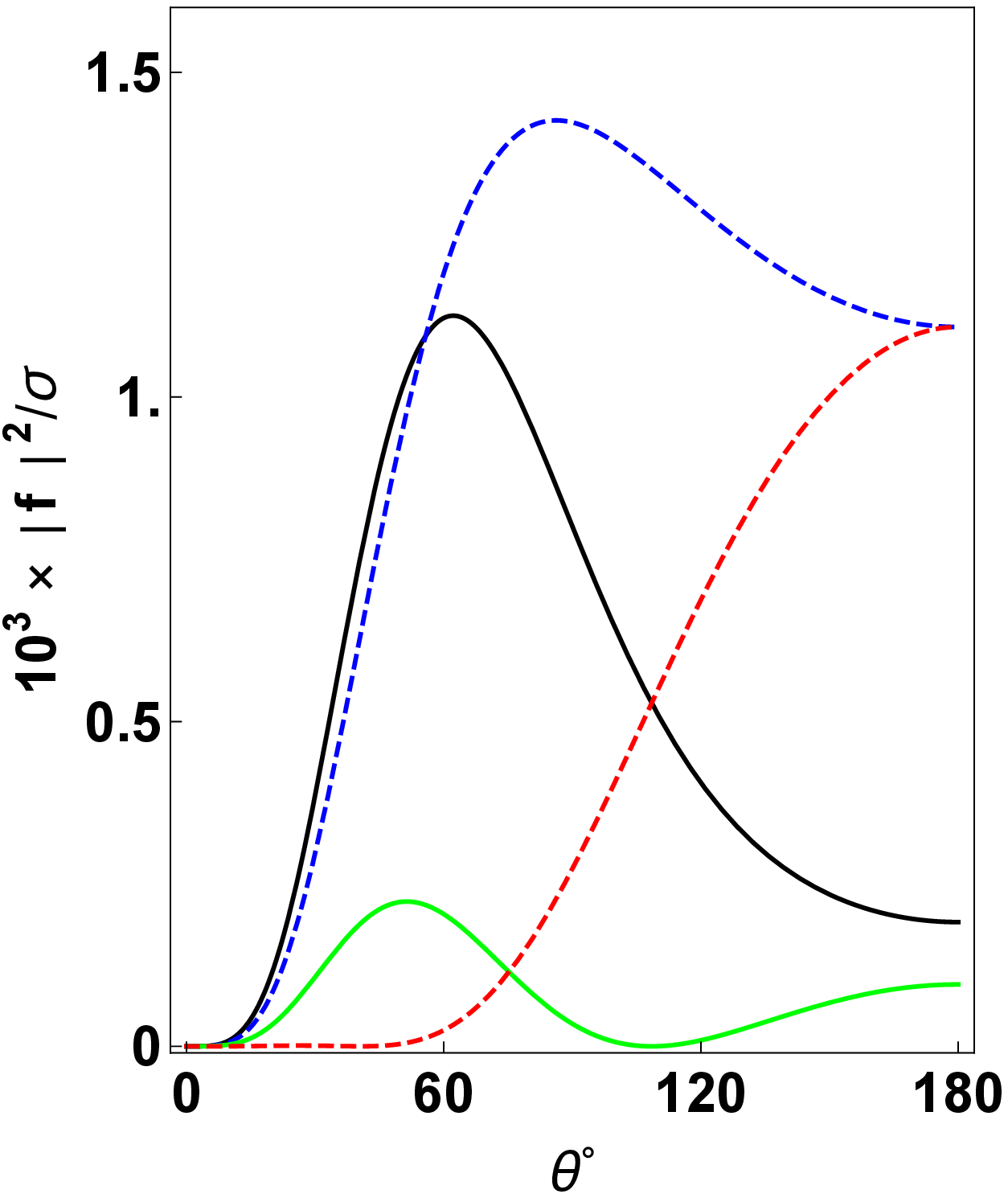}~~~
        \includegraphics[scale=.43]{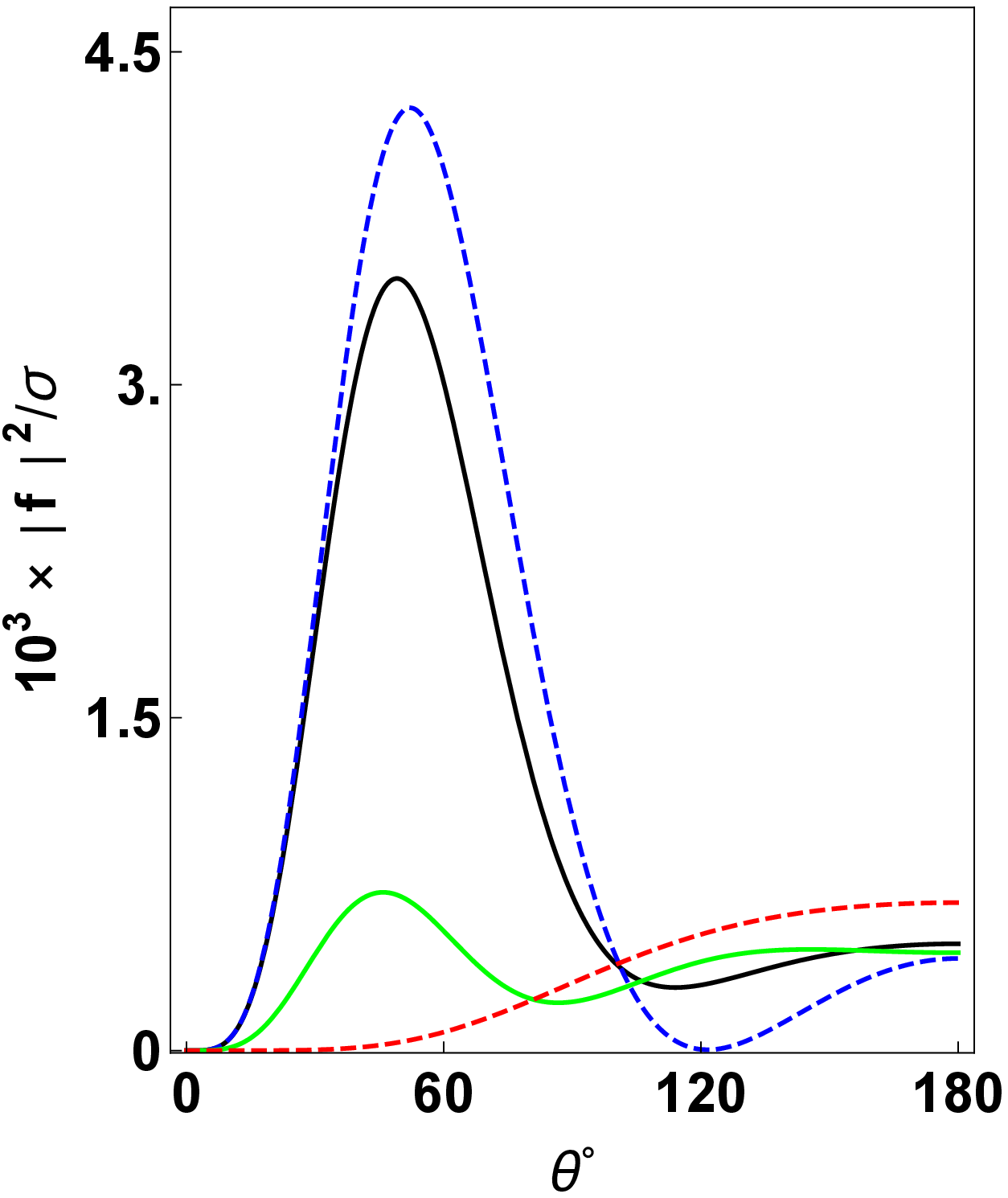}
        \caption{Plots of $|f(\boldsymbol{k}',\boldsymbol{k})|^2/\sigma$  as functions of $\theta$ for the Gaussian bump \eqref{eq47} with a line defect at $x=-3\sigma$ (on the left),  $x=0$ (in the middle), and $x=3\sigma$ (on the right) for $\theta_0=0^\circ$, $\eta=0.1$, $\sigma\fz_1=k\sigma=1$, and different values of $\lambda_1$ and $\lambda_2$, namely $\lambda_1=-\lambda_2=1/2$ (black), $\lambda_1=0$ and $\lambda_2=-1/2$ (dashed blue), $\lambda_1=1/2$ and $\lambda_2=0$ (green), and $\lambda_1=\lambda_2=1/2$ (dashed red).}
        \label{fig3}
    \end{center}
    \end{figure}

\begin{figure}
    \begin{center}
        \includegraphics[scale=.43]{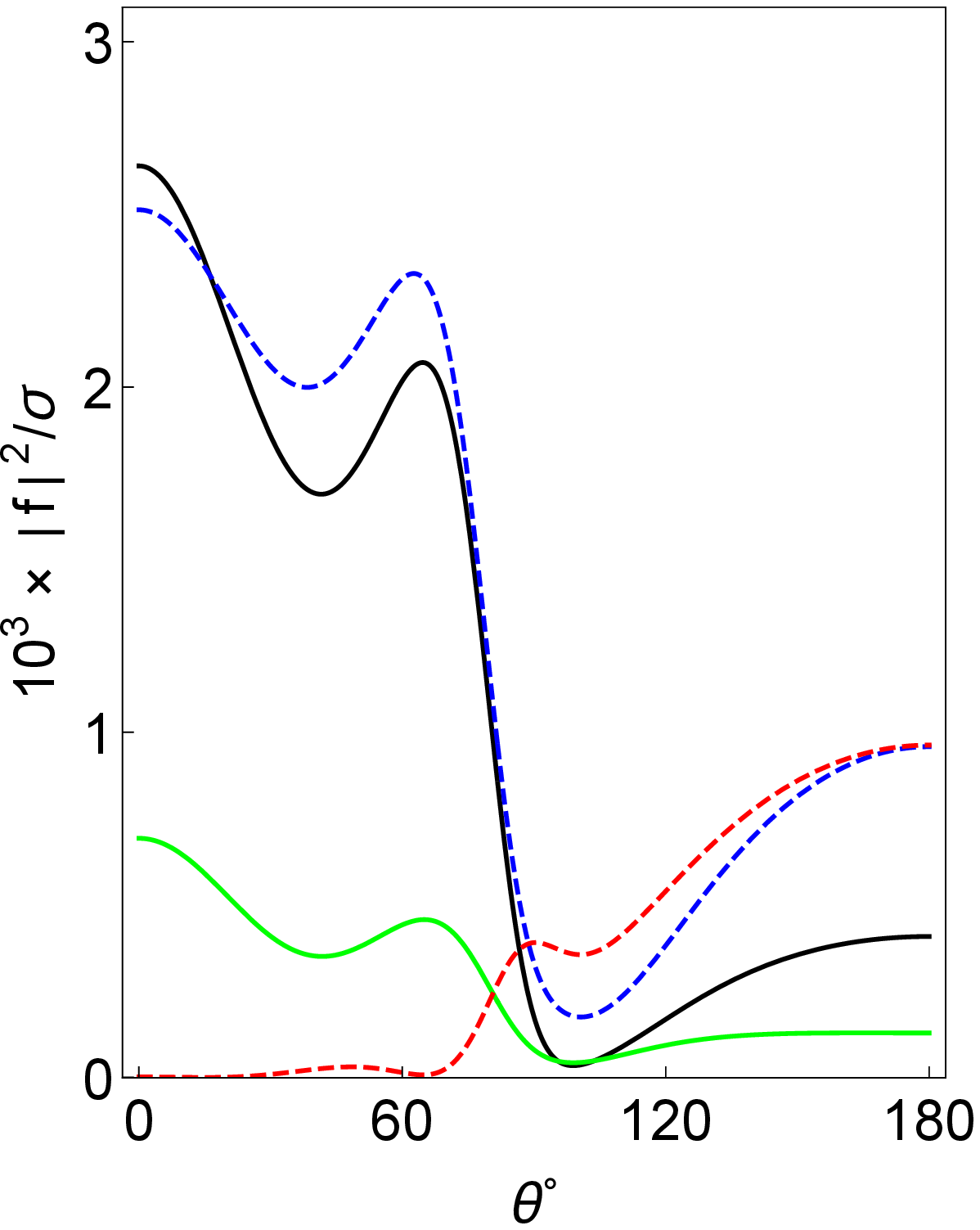}~~~
        \includegraphics[scale=.43]{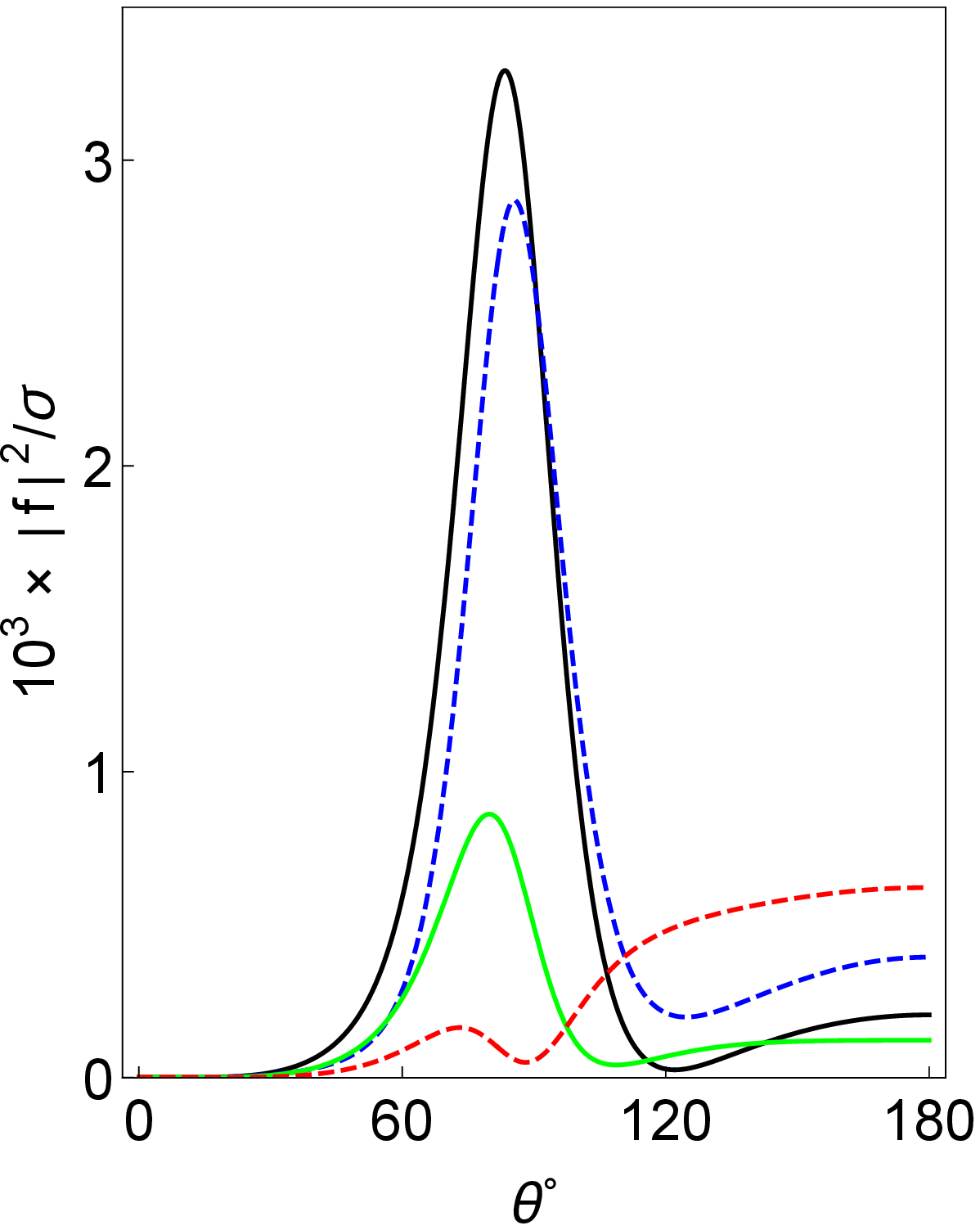}~~~
        \includegraphics[scale=.445]{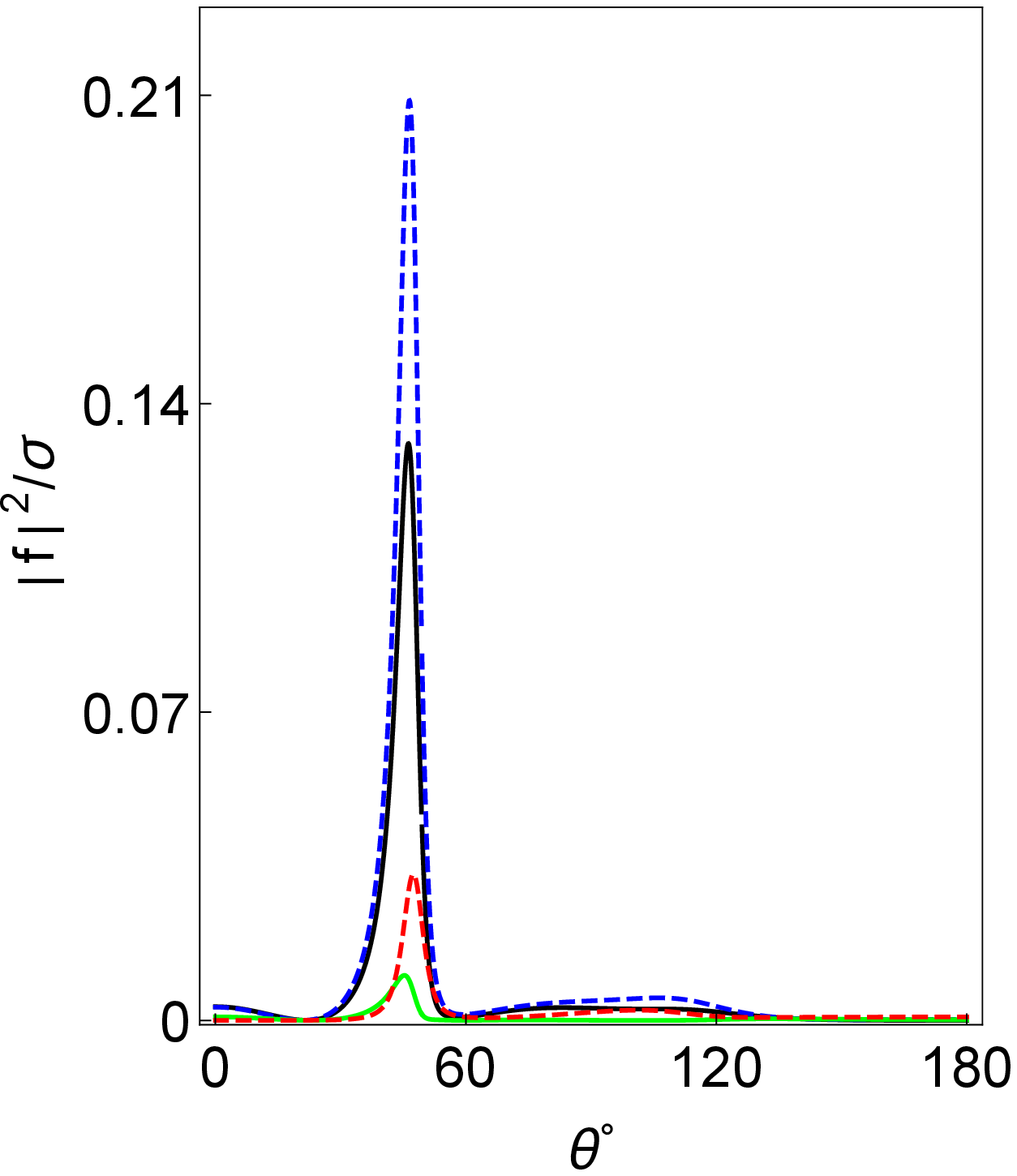}
        \caption{Plots of $|f(\boldsymbol{k}',\boldsymbol{k})|^2/\sigma$  as functions of $\theta$ for the Gaussian bump \eqref{eq47} with two line defects at $x=-3\sigma$ and $x=0$ (on the left),  $x=0$ and $x=3\sigma$ (in the middle), and $x=\pm 3\sigma$ (on the right) for $\theta_0=0^\circ$, $\eta=0.1$, $\sigma\fz_1=k\sigma=1$, and different values of $\lambda_1$ and $\lambda_2$, namely $\lambda_1=-\lambda_2=1/2$ (black), $\lambda_1=0$ and $\lambda_2=-1/2$ (dashed blue), $\lambda_1=1/2$ and $\lambda_2=0$ (green), and $\lambda_1=\lambda_2=1/2$ (dashed red).}
        \label{fig4}
    \end{center}
    \end{figure}

Figures~\ref{fig3} and \ref{fig4} show the plots of $|\ff(\bk',\bk)|^2$ as a function of the scattering angle $\theta$ for a Gaussian bump in the presence of one or two line defects with different values of the curvature coefficients $\lambda_1$ and $\lambda_2$. Here we have set $\fK=k\sigma=1$.
The behavior of the differential cross section depicted in Figs.~\ref{fig3} and \ref{fig4} is consistent with that of Figs.~\ref{fig1} and \ref{fig2}; for the case of a single line defect it is smaller when the defect passes through the center of the bump, and for the case of two line defects it is much larger when the defects are placed symmetrically about the bump. According to Figs.~\ref{fig3} and \ref{fig4}, different choices for the curvature coefficients lead to differential cross sections with completely different characteristics. This should facilitate the experimental determination of these coefficients using the scattering data.

\begin{figure}[!ht]
    \begin{center}
        \includegraphics[scale=.5]{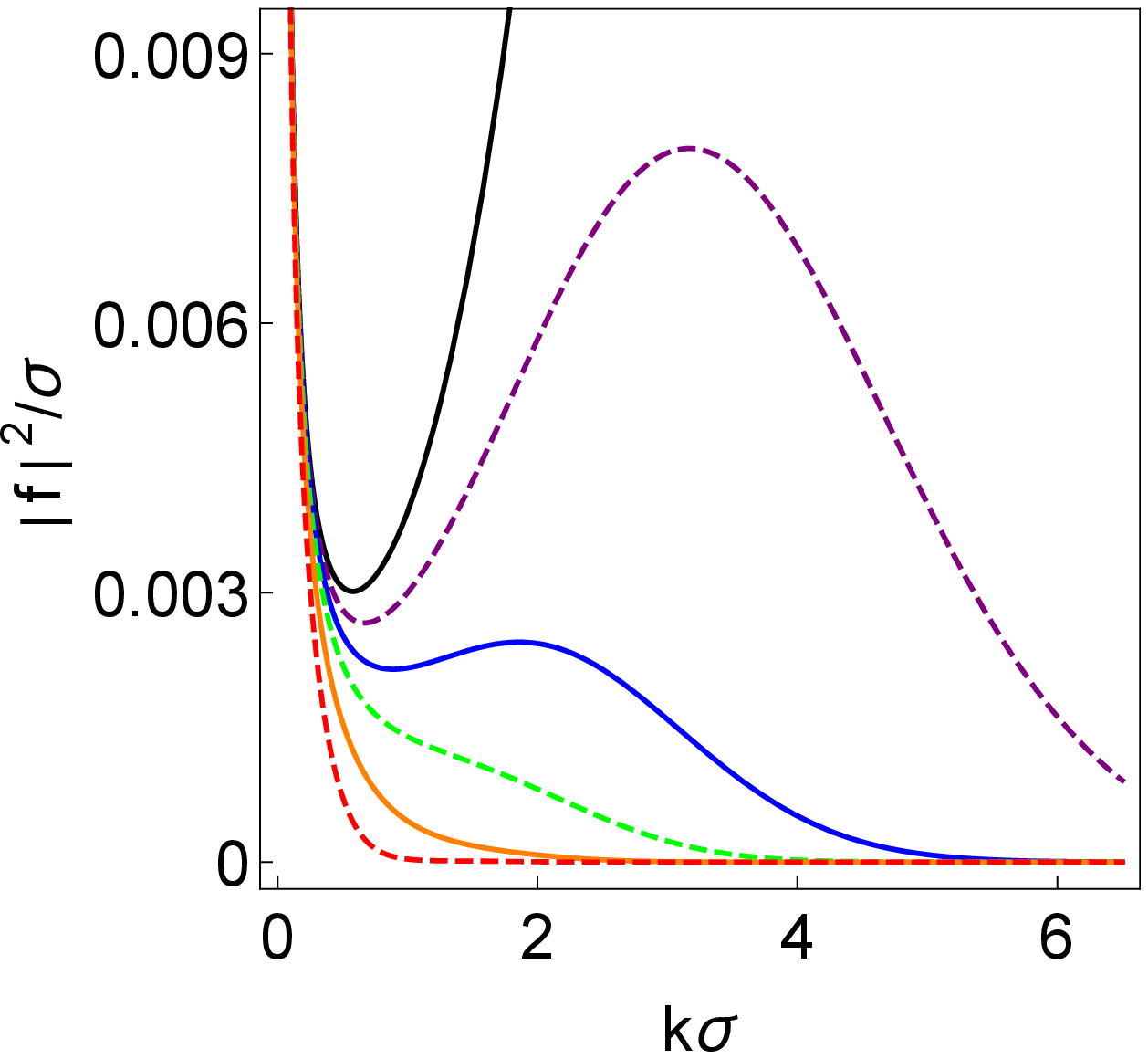}~~~~~~
        \includegraphics[scale=.54]{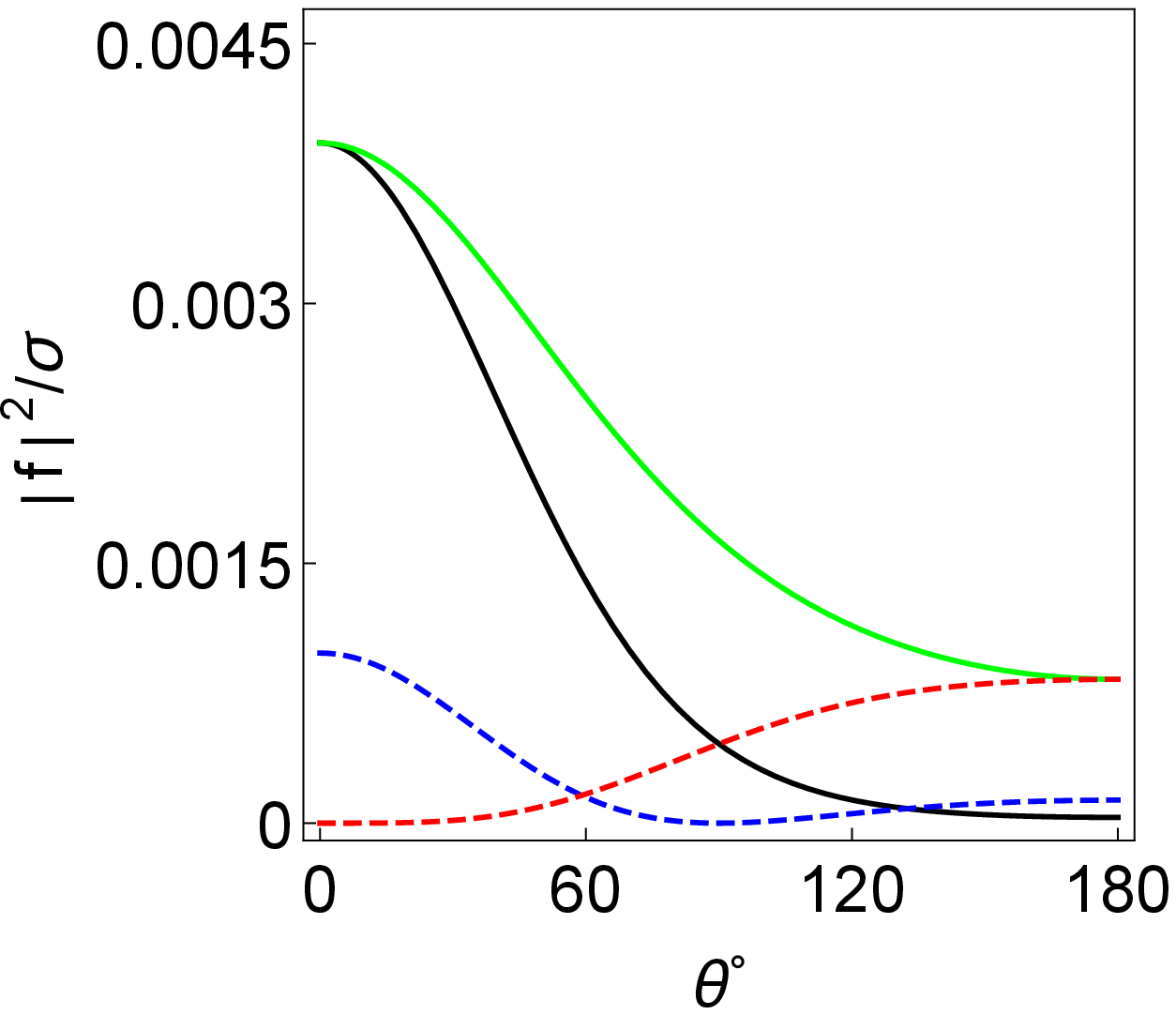}
        \caption{Plots of $|f(\boldsymbol{k}',\boldsymbol{k})|^2/\sigma$ as functions of $\fK=k\sigma$ (on the left) and $\theta$ (on the right) for the Gaussian bump \eqref{eq47} in the absence of the line defects with the same values of the physical parameters as in Figs~\ref{fig1} and~\ref{fig2}. The graphs in the left panel correspond to the scattering angles:  $\theta=5^\circ$ (black), $\theta=30^\circ$ (dashed purple), $45^\circ$ (blue), $60^\circ$ (dashed green), $90^\circ$ (orange), and $175^\circ$ (dashed red). Those in the right panel correspond to the curvature coefficients: $\lambda_1=-\lambda_2=1/2$ (black), $\lambda_1=0$ and $\lambda_2=-1/2$ (dashed blue), $\lambda_1=1/2$ and $\lambda_2=0$ (green), and $\lambda_1=\lambda_2=1/2$ (dashed red).}
    \label{fig5}
    \end{center}
\end{figure}

To decide if the presence of the line defects enhances the geometric scattering effects,
 we have also plotted in Fig.~\ref{fig5} the graphs of the differential cross section $|\ff(\bk',\bk)|^2$ in the absence of the line defects for the same parameters as those used Figs.~\ref{fig1} -- \ref{fig4}.

Comparing the graphs given in Figs.~\ref{fig1} and \ref{fig2} with the graph on the left in Fig.~\ref{fig5}, we see that the presence of line defects enhances the geometric scattering effects considerably. This is particularly strong when  the center of the bump is placed between two lines. For the case that these line defects are located at the distance $3\sigma$ from the center of the bump, the peak of the differential cross section for $\theta_0=30^\circ$ is larger than its peak in the absence of the line defects by about two orders of magnitude. 

Inclusion of point defects also amplify the scattering of waves by the surface \cite{ap-2019}. Note however that their presence contributes to the differential cross section at every scattering angle $\theta$, while line defects only produce reflected and transmitted rays at $\theta=\theta_0$ and $180^\circ-\theta_0$. Therefore the behavior of the differential cross section for the scattering angles other than $\theta_0$ and $180^\circ-\theta_0$ reflects the influence of the line defects on the geometric scattering due to the surface.

\section{Concluding remarks}

Quantization of a classical nonrelativistic particle moving in a curved surface that is embedded in the three-dimensional Euclidean space does not yield a unique quantum system. The non-uniqueness of the quantum system is characterized by the choice of two free parameters. These enter the expression for the Hamiltonian operator as coefficients of terms involving the Gaussian and mean curvatures of the surface. For an asymptotically flat surface, these curvature terms contribute to the scattering amplitude of the particle. Realizing this phenomenon in a dilute electron gas formed on a Gaussian bump requires dealing with the problem of defects. A previous study shows that point defects amplify the geometric scattering effects of the surface \cite{ap-2019}. In the present article, we have examined the influence of line defects. This is motivated by the idea that placing the center of the bump between a pair of parallel line defects can produce an effective resonator capable of achieving much larger amplification of the geometric scattering effects. To examine the feasibility of this idea, we have addressed the scattering problem for the general case where the surface includes $N$ line defects. 

Line defects placed on a Euclidean plane scatter the particle along two specific directions; the scattering amplitude vanishes expect for the scattering angles $\theta_0$ and $180^\circ-\theta_0$, where $\theta_0$ is the angle of incidence. For a curved embedded surface $S$, the scattering amplitude is a smooth nonzero function of the scattering angle $\theta$. This shows that for values of $\theta$ different from $\theta_0$ and $180^\circ-\theta_0$, the scattering phenomenon stems from the nontrivial geometry of $S$. A detailed examination of the scattering cross section for this system provides strong support for our expectation regarding the use of line defects for the purpose of amplifying the geometric scattering effects.

\section*{Appendix~A: Derivation of Eq.~(\ref{eq20})}

The scattering solution (\ref{eq12}) of the Schr\"odinger equation (\ref{eq2}) has the form
    \be
    \psi_0(\bx)=\psi_{\rm inc}(\bx)+\psi_{\rm scatt}(\bx),
    \label{app-e1}
    \ee
where $\psi_{\rm inc}(\bx)$ and $\psi_{\rm scatt}(\bx)$ respectively represent the incident and scattered waves and have the form $\psi_{\rm inc}(\bx):=e^{i\bk\cdot\bx}/2\pi$ and
    \be
    \psi_{\rm scatt}(\bx):=\frac{-i}{2\pi}
    \sum_{m,n=1}^N e^{i k_x a_m}A^{-1}_{mn}\,
    e^{i(k_x |x-a_n|+k_y y)}.
    \label{app-e2}
    \ee
In view of (\ref{asymp}) and (\ref{app-e1}),
    \be
    \psi_{\rm scatt}(\bx)\to \frac{e^{ik r}}{2\pi\sqrt{r}}\,\ff_0(\bk',\bk)~~~{\rm for}~~~r\to\infty.
    \label{asymp2}
    \ee
This shows that in order to compute the scattering amplitude, we should determine the large-$r$ behavior of the the right-hand side of (\ref{app-e2}). First, we express this equation as
    \be
    \psi_{\rm scatt}(\bx)=\frac{e^{ik_y y}}{2 \pi} \sum_{n=1}^N
    \Big[ \ft^+_{n}\ \Theta (x-a_n)e^{ik_x x} + \ft^-_{n}\ \Theta (a_n-x)e^{-ik_x x}\Big],
    \label{eq13}
    \ee
where
    \begin{align}
    &\ft^\pm_{n}:=-i\sum_{m=1}^N A^{-1}_{mn} \ e^{i k_x (a_m\mp a_n)},
    &&\Theta(x):=\left\{\begin{array}{ccc}
    0 & {\rm for} & x<0,\\
    1 & {\rm for} & x\geq 0.\end{array}\right.
    \label{eq14}
    \end{align}

For the scattering setup we consider, the source of the incident way lies at $x=-\infty$. This implies that $k_x>0$ and the incidence angle $\theta_0$ takes values in the interval $(-\frac{\pi}{2},\frac{\pi}{2})$. Because the angular position of the detector is arbitrary, we take the scattering angle $\theta$ to range over the interval $[-\frac{\pi}{2},\frac{3\pi}{2})$. Next, we introduce the notation:
    \begin{align}
    &\theta^+:=\theta~~~~~~~~~{\rm for}~~~\mbox{$\theta\in(-\frac{\pi}{2},\frac{\pi}{2})$},\nn\\
    &\theta^{-}:=\pi - \theta~~~{\rm for}~~~\mbox{$\theta \in (\frac{\pi}{2},\frac{3\pi}{2})$},\nn
    \end{align}
and employ the analysis presented in Appendix~A of Ref.~\cite{pra-2016} to establish the identity:
    \begin{equation}
    \label{eq18}
    e^{ik_y y}e^{\pm ik_x x} 
    \rightarrow\sqrt{\frac{2\pi }{k r}} \ \Big[e^{i(kr-\frac{\pi}{4})}\delta(\theta_0-\theta^{\pm})+e^{-i(kr-\frac{\pi}{4})}    \delta(\theta_0-\theta^{\pm}+ \pi)\Big]~~{\rm as}~~r\to\infty.
    \end{equation}
With the help of this relation and Eqs.~(\ref{eq13}) and (\ref{eq14}), we obatin (\ref{asymp2}) with $\ff_0(\bk',\bk)$ given by (\ref{eq20}).

\subsection*{Appendix B: Formulas for $I_{mn}, J_{mn}$, and $I_{mm'nn'}$}

The following are the formulas we have obtained for $I_{mn}, J_{mn}$, and $I_{mm'nn'}$ by performing the integrals in (\ref{eq32}) -- (\ref{eq34}). Here $\alpha_n:=a_n/\sigma$, and $\text{Erf}[x]$ and $\text{Erfc}[x]$ are respectively the error and complementary error functions.\footnote{By definition, $\text{Erfc}[x]:=1-\text{Erf}[x]$.}
    \begin{align*}
    I_{mn}=&\frac{1}{8} \eta  e^{-s \fK (-i {\alpha_m}+i \alpha_n+s \fK)} \Bigg\{\sqrt{\pi }  s \fK e^{(s \fK+i \alpha_n)^2} \Big[-2 i \alpha_n^2 \lambda_2-2 \alpha_n (\lambda_2-2) s \fK+i \left(8 \lambda_1+\lambda_2-2 \lambda_2 s^2 \fK^2\right)\Big]\nn\\
    &+2 \pi  \text{Erf}[\alpha_n-i s \fK] \Big[2 \lambda_2+\lambda_2 s^4 \fK^4+\fK^2 (4 \lambda_1 s^2-1)\Big]-2 \pi   \text{Erfc}[\alpha_n] (\fK^2-2 \lambda_2) e^{s \fK (s \fK+2 i \alpha_n)}\nn\\
    &+2 \pi   \Big[2 \lambda_2+\lambda_2 s^4 \fK^4+\fK^2 (4 \lambda_1 s^2-1)\Big]\Bigg\}+\cO(\eta^2),\\
    J_{mn}=&\frac{1}{8} \eta  e^{i {\alpha_m} s \fK} \sqrt{\pi }  \Bigg\{2 e^{-s \fK (i {\alpha_n}+s \fK)} \sqrt{\pi }
    \left[\fK^2 \left(-1+4 s^2 \lambda_1\right)+2 \lambda_2+s^4 \fK^4 \lambda_2\right] \text{Erfc}[{\alpha_n}-i s   \fK]\\
    &-2 e^{i {\alpha_n} s \fK} \sqrt{\pi } (\fK^2-2 \lambda_2) \text{Erf}[{\alpha_n}]+e^{-{\alpha_n} ({\alpha_n}-i s    \fK)} \Big[(-2 e^{{\alpha_n}^2} \sqrt{\pi } (\fK^2-2 \text{$\lambda
        $2})-i s \fK (-4+8 \lambda_1\\
    &-2 i {\alpha_n} s \fK (-2+\lambda_2)+\lambda_2+2 s^2 \fK^2 \lambda_2-2 {\alpha_n}^2 (4+\lambda_2))\Big]\Bigg\}+\cO(\eta^2),\\
    I_{mm'nn'}=&g_{mm'nn'}+\Theta(m-n)h_{mm'nn'}+\Theta(n-m)k_{mm'nn'}+l_{mm'nn'},
    \end{align*}
where
    \begin{equation}
    g_{mm'nn'}:=\begin{cases}
    q_{mm'nn'}&\text{for}\qquad m=n,\\
    s_{mm'nn'}&\text{for} \qquad m > n,\\
    t_{mm'nn'}&\text{for}\qquad m<n,
    \end{cases}\nn
    \end{equation}
    \begin{align*}
    q_{mm'nn'}:=&\frac{1}{4} \sqrt{\pi } \eta  s \fK [e^{i s \fK (\alpha_m'+{\alpha_n'})}] e^{-\alpha_m^2-i s \fK   (\alpha_m-{\alpha_n})-{\alpha_n}^2} \Bigg[e^{\alpha_m^2} \left(\sqrt{\pi } e^{{\alpha_n}^2} s \fK (\text{Erfc}[{\alpha_n})-2]-2 i {\alpha_n}^2+2 {\alpha_n} s \fK+i\right)\\
    &-i e^{{\alpha_n}^2+2 i s \fK (\alpha_m-{\alpha_n})} \left(-i s \fK \left(\sqrt{\pi } e^{\alpha_m^2} \text{Erfc}[\alpha_m]+2 \alpha_m\right)+2 \alpha_m^2-1\right)\Bigg]+\cO(\eta^2),
    \end{align*}
    \begin{align*}
    s_{mm'nn'}:=&\frac{1}{4} \sqrt{\pi } \eta  s \fK [e^{i s \fK (\alpha_m'+{\alpha_n'})}]e^{-\alpha_m^2-s \fK (s \fK+i ({\alpha_m}+{\alpha_n}))-{\alpha_n}^2} \Bigg[e^{{\alpha_n}^2} s \fK \Bigg(\sqrt{\pi } e^{{\alpha_m}^2} \Big(\text{Erfc}[{\alpha_m}-i s \fK]+\text{Erf}[{\alpha_n}-i s \fK]\\
&+(\text{Erfc}[{\alpha_n}]-2) e^{s \fK (s \fK+2 i {\alpha_n})}-1\Big)-(\sqrt{\pi } e^{{\alpha_m}^2} \text{Erfc}[{\alpha_m}]+2 {\alpha_m}) e^{s \fK (s \fK+2 i {\alpha_m})}\Bigg)\\
&+2 (-2 i {\alpha_n}^2+{\alpha_n} s \fK+i) e^{{\alpha_m}^2+s \fK (s \fK+2 i {\alpha_n})}\Bigg]+\cO(\eta^2),
    \end{align*}
    \begin{align*}
    t_{mm'nn'}:=&\frac{1}{4} \sqrt{\pi } \eta  s \fK e^{i s \fK (\alpha_m'+{\alpha_n'})} e^{-{\alpha_m}^2-s \fK (s \fK+i ({\alpha_m}+{\alpha_n}))-{\alpha_n}^2} \Bigg[s \fK \left(\sqrt{\pi } e^{{\alpha_m}^2} (\text{Erfc}[{\alpha_m}]-2)+2 {\alpha_m}\right) e^{{\alpha_n}^2+s \fK (s \fK+2 i {\alpha_n})}\\
&+e^{{\alpha_m} ({\alpha_m}+2 i s \fK)} \Bigg(\sqrt{\pi } s \fK e^{{\alpha_n} ({\alpha_n}+2 i s \fK)} (\text{Erf}[{\alpha_m}+i s \fK]-\text{Erf}[{\alpha_n}+i s \fK])\\
&-e^{s^2 \fK^2} \Big(\sqrt{\pi } e^{{\alpha_n}^2} s \fK \text{Erfc}[{\alpha_n}]+4 i {\alpha_n}^2+2 {\alpha_n} s \fK-2 i\Big)\Bigg)\Bigg]+\cO(\eta^2),
    \end{align*}
    \begin{align*}
    h_{mm'nn'}:=&\frac{1}{16} \eta e^{i s \fK (\alpha_m'+{\alpha_n'})}\Bigg\{12 \pi  \eta  (\text{Erf}[{\alpha_n}]-\text{Erf}[{\alpha_m}]) \left(2 \lambda_2+\left(s^2-1\right) \fK^2\right) e^{i s \fK ({\alpha_m}-{\alpha_n})}\\
&+4 \pi  \eta  e^{-i s \fK ({\alpha_m}+{\alpha_n}-i s \fK)} \text{Erf}[{\alpha_m}-i s \fK] \left(4 \lambda_1+6 \lambda_2+3 \lambda_2 s^4 \fK^4+\fK^2 \left((4 \lambda_1+3) s^2-3\right)\right)\\
&-4\pi  \eta  e^{-s \fK (s \fK+i ({\alpha_m}+{\alpha_n}))} \text{Erf}[{\alpha_n}-i s \fK] \left(4 \lambda_1+6 \lambda_2+3 \lambda_2 s^4 \fK^4+\fK^2 \left((4 \lambda_1+3) s^2-3\right)\right)\\
&+ \sqrt{\pi } e^{-{\alpha_m}^2-{\alpha_n}^2} \Bigg[2 e^{-i s \fK ({\alpha_m}+{\alpha_n})} \Bigg(i e^{{\alpha_m}^2+2 i {\alpha_n} s \fK} \Big(6 {\alpha_n}^2 (\lambda_2+8) s \fK+i {\alpha_n} \left(8 \lambda_1+9 \lambda_2+6 \lambda_2 s^2 \fK^2\right)\\
&-s \fK \left(8 \lambda_1+3 \lambda_2+6 \lambda_2 s^2 \fK^2\right)\Big)+e^{2 i {\alpha_m} s \fK} \Big(3 e^{{\alpha_m}^2} {\alpha_n} (8 \lambda_1+3 \lambda_2)+e^{{\alpha_n}^2} \big(3 \lambda_2 s \fK (-2 i {\alpha_m}^2\\
&+2 {\alpha_m} s \fK+2 i s^2 \fK^2+i)-16 {\alpha_m} \lambda_1+8 i \lambda_1 s \fK\big)\Big)\Bigg)-24 i e^{{\alpha_m}^2} {\alpha_n}^3 \lambda_2 \sin (s \fK ({\alpha_m}-{\alpha_n}))\Bigg]\Bigg\}+\cO(\eta^2),
    \end{align*}
    \begin{align*}
    k_{mm'nn'}:=&\frac{1}{8} \sqrt{\pi } \eta e^{i s \fK (\alpha_m'+{\alpha_n'})} \Bigg\{e^{-{\alpha_m}^2-{\alpha_n}^2-i ({\alpha_m}-{\alpha_n}) s \fK} \Big(-{\alpha_m} e^{{\alpha_n}^2} \left(-8 \lambda_1+\left(-3+2 {\alpha_m}^2\right) \lambda_2\right)\\
    &+{\alpha_n} e^{{\alpha_m}^2} \left(-8 \lambda_1+\left(-3+2 {\alpha_n}^2\right) \lambda_2\right)\Big)+ e^{-{\alpha_m}^2-{\alpha_n}^2-i ({\alpha_m}+{\alpha_n}) s \fK} \Big[e^{{\alpha_m} ({\alpha_m}+2 i s \fK)} \Big(8 {\alpha_n} \lambda_1-2 {\alpha_n}^3 \lambda_2\\
    &+{\alpha_n} \left(3+2 s^2 \fK^2\right) \lambda_2+2 i {\alpha_n}^2 s \fK (8+\lambda_2)-i s \fK \left(8 \lambda_1+\lambda_2+2 s^2 \fK^2 \lambda_2\right)\Big)\\
    &+e^{{\alpha_n} ({\alpha_n}+2 i s \fK)} \Big(2 {\alpha_m}^3 \lambda_2-2 i {\alpha_m}^2 s \fK \lambda_2+i s \fK \left(8 \lambda_1+\lambda_2+2 s^2 \fK^2 \lambda_2\right)-{\alpha_m} \left(8 \lambda_1+\left(3+2 s^2 \fK^2\right) \lambda_2\right)\Big)\Big]\\
    &+2 e^{-i ({\alpha_m}-{\alpha_n}) s \fK} \sqrt{\pi } \left(\left(s^2-1\right) \fK^2+2 \lambda_2\right) \text{Erf}[{\alpha_m}]-2 e^{-i ({\alpha_m}-{\alpha_n}) s \fK} \sqrt{\pi } \left(\left(s^2-1\right) \fK^2+2 \lambda_2\right) \text{Erf}[{\alpha_n}]\\
    &-2 e^{i s \fK ({\alpha_m}+{\alpha_n}+i s \fK)} \sqrt{\pi } \left(x^2 \left(-1+s^2 (1+4 \lambda_1)\right)+2 \lambda_2+s^4 \fK^4 \lambda_2\right) \text{Erf}[{\alpha_m}+i s \fK]\\
    & + 2 e^{i s \fK ({\alpha_m}+{\alpha_n}+i s \fK)} \sqrt{\pi } \left(\fK^2 \left(-1+s^2 (1+4 \lambda_1)\right)+2 \lambda_2+s^4 \fK^4 \lambda_2\right) \text{Erf}[{\alpha_n}+i s \fK]\Bigg\}+\cO(\eta^2),
    \end{align*}
    \begin{align*}
    l_{mm'nn'}:=&\frac{1}{8}\eta e^{i s \fK (\alpha_m'+{\alpha_n'})} e^{-{\alpha_n}^2+i s \fK ({\alpha_n}-{\alpha_m})} \Bigg[2 \pi  e^{{\alpha_n}^2}  \left(2 \lambda_2+\left(s^2-1\right) \fK^2\right) \left(\text{Erf}[{\alpha_n}]+\text{Erfc}[{\alpha_n}] e^{2 i s \fK ({\alpha_m}-{\alpha_n})}\right)\\
    &+\sqrt{\pi }  \Bigg({\alpha_n} \left(\left(2 {\alpha_n}^2-3\right) \lambda_2-8 \lambda_1\right) e^{2 i s \fK ({\alpha_m}-{\alpha_n})}+2 \sqrt{\pi } e^{{\alpha_n}^2} \left(2 \lambda_2+\left(s^2-1\right) \fK^2\right)\\
    &-2 {\alpha_n}^3 \lambda_2+8 {\alpha_n} \lambda_1+3 {\alpha_n} \lambda_2\Bigg)\Bigg]+\cO(\eta^2).
    \end{align*}
    
\noindent{\bf Acknowledgements.} This work has been supported by the Scientific  and Technological Research Council of Turkey (T\"UB{$\dot{\rm I}$}TAK) in the framework of the Project
No.~117F108 and by the Turkish Academy of Sciences (T\"UBA).

\ed